\documentclass[nofootinbib, twocolumn,
aps,
superscriptaddress,
pra
]{revtex4-1}

\usepackage[colorlinks=true,urlcolor=blue,citecolor=blue,linkcolor=red]{hyperref}
\usepackage{graphicx}
\usepackage{wrapfig}
\usepackage{booktabs}
\usepackage{amsmath}
\usepackage{amssymb}
\usepackage{braket}
\usepackage[dvipsnames]{xcolor}
\usepackage[normalem]{ulem}
\usepackage{siunitx}
\usepackage[T1]{fontenc}
\usepackage{hyperref}
\usepackage{bbold}

\begin{document}
	\title{Quantum spacetime from constraints: wave equations and fields}

	\author{Tommaso Favalli}\email{tommaso.favalli@units.it}
	\affiliation{University of Trieste, Strada Costiera 11, I-34151 Trieste, Italy}

	\begin{abstract}
In previous works, we showed that both time and space can emerge from entanglement within a globally constrained quantum Universe, with no background coordinates. By extending the Page and Wootters quantum time formalism to include both quantum clocks and rods, and imposing global constraints on total energy and momentum, we constructed a fully relational model of quantum spacetime. Here we take a further step: working in 1+1 dimensions, we show that the standard wave equations governing quantum particles—the Schrödinger, Klein-Gordon and Dirac equations—emerge naturally from this framework. The solutions of the equations are derived directly from the constraints, without assuming any external spacetime structure. The second quantization formalism is also implemented and discussed. Our results provide further support for the idea that quantum dynamics in spacetime may emerge from entanglement and constraints.
	\end{abstract}
	
	\maketitle

\section{Introduction}
	
In the quantum gravity literature, quantum reference frames have been suggested to play a key role in formulating a consistent and workable quantum theory of gravity \cite{dewitt,afundamental,QG1,QG2}. This idea stems from the observation that, in the absence of an external classical background, notions such as time and space must be defined relationally with respect to other quantum degrees of freedom. This viewpoint is deeply rooted in canonical approaches to quantum gravity, in particular in quantum geometrodynamics, where the dynamics of the Universe is encoded in global constraints. These include the Hamiltonian constraint and the diffeomorphism constraints, which are implemented through the Wheeler--DeWitt equation \cite{dewitt,isham}. In this framework, the global state of the Universe is stationary, and temporal evolution can only be recovered in a relational or approximate manner, giving rise to the so-called \lq\lq problem of time\rq\rq.

The Page and Wootters (PaW) mechanism \cite{pagewootters,wootters,pagearrow} provides a concrete realization of this idea. By partitioning the total Hilbert space into subsystems and identifying one of them as a quantum clock, it becomes possible to recover an effective Schrödinger evolution for the remaining degrees of freedom, conditioned on the clock observable. In this way, time is not introduced as an external parameter, but rather emerges from correlations within a globally constrained quantum state. This framework has been further developed and clarified over the years, addressing conceptual and technical issues related to the definition of time observables, conditional states, and the role of entanglement in generating dynamical evolution (see, e.g., Refs.~\cite{lloydmaccone,vedral,vedraltemperature,chapter2,nostro,nostro2,nostro3,librotommi,nostro4,interacting,timedilation,simile,simile2,review,review2,scalarparticles,dirac,foti,brukner,esp1,esp2}).

In parallel, significant progress has been made in the formulation of quantum reference frames (QRFs), where physical laws are expressed relative to quantum systems rather than classical observers. These approaches have demonstrated that quantum theory admits a consistent, observer-dependent formulation, in which different choices of reference frame correspond to different but equivalent descriptions of the same physical situation~\cite{change1,change2,change3,change4,change5,change6,change7,change8,change9,hoehn1,hoehn2,hoehn3}. More recent developments have extended these ideas beyond purely kinematical settings, exploring relational descriptions that incorporate aspects of spacetime and address situations in which both temporal and spatial notions are defined relationally (see Refs.~\cite{giacomini,nuovo1,nuovo2,nuovo3}).

Related lines of research have also explored the possibility of formulating quantum mechanics directly in terms of spacetime or global constraints, without relying on a preferred external time parameter. In particular, spacetime quantum formulations and action-based approaches have been developed, including treatments with dynamical foliations, path-integral constructions, and extensions to both bosonic and fermionic systems \cite{dias1,dias2,dias3,dias4}. Other works have investigated the emergence of quantum space and time, as well as relativistic quantum dynamics, from closely related conceptual perspectives \cite{nuovo32}. These approaches collectively suggest that the standard dynamical laws of quantum theory may arise from deeper, constraint-based structures.

In previous works \cite{nostro3,librotommi}, we contributed to this line of research by extending the PaW formalism to include both temporal and spatial reference frames. In that framework, time and position emerge from entanglement in a globally \lq\lq timeless\rq\rq\ and \lq\lq positionless\rq\rq\ Universe, described by a quantum state subject to global constraints on total energy and momentum. By introducing quantum clocks and rods as internal subsystems, we showed that the conditional states of a physical system evolve coherently in both time and space, leading to a fully relational description of non-relativistic emergent spacetime, formulated without any external background geometry.

The present work builds on this framework and further develops its implications. While quantum reference frame approaches primarily focus on transformations between different quantum perspectives, and spacetime quantum approaches reformulate dynamics in terms of spacetime structures or action-based formulations, here we adopt a complementary viewpoint. Starting from the relational PaW mechanism, we introduce both quantum clocks and quantum rods as internal subsystems and show how standard quantum wave equations emerge directly from global constraints.

In particular, we show how the Schrödinger, Klein--Gordon, and Dirac equations, together with their corresponding solutions, can be recovered within a relational setting in which temporal and spatial structures are encoded in correlations between subsystems. Rather than constituting first-principles derivations, these results show how the equations can be embedded in the constrained structure of the total quantum state.

To this end, we consider a Universe composed of a clock subsystem $C$ and two quantum particles. The clock provides the temporal reference, while the two particles play distinct roles: one acts as a spatial reference frame $R$, and the other as the physical system $S$ under investigation. In general, both $R$ and $S$ evolve according to the global constraints. However, under the assumption that the kinetic energy of the reference particle $R$ is negligible—e.g., by considering a sufficiently large mass $M$—one can effectively extract the dynamics of $S$ alone. In the Schrödinger case, we also analyze situations in which this approximation is relaxed, allowing for a non-negligible contribution of $R$ and for interactions between $R$ and $S$.

Furthermore, we introduce a second-quantized formulation within this relational framework. This extension is intended to illustrate how standard field-theoretic structures can be consistently embedded in the present setting. In this work, we restrict the analysis to the sector effectively describing a single excitation, using the second-quantized formalism primarily as a representational tool rather than to address the full many-particle regime. A more complete treatment of genuinely many-particle aspects is left for future work. In this sense, the present framework accommodates both effective single-particle dynamics and field-theoretic structures within a unified relational setting based on global constraints.

The paper is organized as follows. Section II reviews the framework of quantum spacetime from global constraints. In Section III, we study the Schrödinger equation, while in Sections IV and V we address the relativistic Klein-Gordon and Dirac cases, respectively. In Section VI we introduce the second quantization formalism and in Section VII we give our conclusions and outlook. Finally, we note that throughout this paper we work in \(1+1\) dimensions (i.e., considering a single spatial coordinate) to facilitate the reading, since the extension to \(3+1\) spacetime is straightforward (requiring independent momentum constraints for each spatial direction) and has already been discussed in Refs.~\cite{nostro3,librotommi}.

\section{Review of quantum spacetime}
In this Section, we briefly review the formalism of emergent quantum spacetime from global constraints, following the framework developed in Refs.~\cite{nostro3,librotommi}. 

\subsection{Subsystems in the Universe}
We consider a closed quantum Universe composed of three non-interacting subsystems: a clock $C$, a reference particle $R$, and a system particle $S$. The total Hilbert space is thus given by $\mathcal{H} = \mathcal{H}_C \otimes \mathcal{H}_R \otimes \mathcal{H}_S$. 

A well-known conceptual issue in the PaW framework concerns the role of the quantum clock. It has been argued that, rather than solving the problem of time, the formalism shifts it to the existence of a suitable clock subsystem, leading to an ambiguity in the resulting notion of dynamics \cite{kuchar}. Various approaches have been proposed either to address this issue (see Refs.~\cite{vedral,review}) or to clarify the conditions under which a consistent notion of relational time can be defined (see Refs.~\cite{librotommi,nostro}).

In this context, it has been suggested that the role of the clock can be played by an environment associated with the system under consideration \cite{nostro2,librotommi}. In realistic physical scenarios, additional degrees of freedom are generically present and can provide a physically motivated candidate for an internal temporal reference. This choice has two main advantages. First, the environment typically has a sufficiently dense spectrum to support a nontrivial and well-defined relational dynamics for the system under consideration (as will be discussed in the next Section). Second, the environment can be taken not to contribute to the relevant spatial momentum, a feature that will be particularly useful in the present analysis, where global energy and momentum constraints are introduced (as we will see in more detail shortly). 

To describe the clock/environment, we consider the Hamiltonian $\hat{H}_C$, which we assume to have a discrete, non-degenerate spectrum with rationally related energy ratios. More specifically, we define a set of $d_{C}$ energy states $\ket{E_i}$ and energy levels $E_i$ with $i=0,1,2,...,d_C -1$ such that $(E_i -E_0)/(E_1 - E_0)=A_i/B_i$, where $A_i$ and $B_i$ are integers with no common factors. We thus obtain ($\hslash=1$): 
\begin{equation}\label{ei}
	E_i = E_0 + r_i \frac{2\pi}{T}
\end{equation}
where $T=2\pi r_1/E_1$, $r_i = r_1 A_i/B_i$ for $i>1$ (with $r_0=0$) and $r_1$ equal to the lowest common multiple of the values of $B_i$. 
In this space we define a continuous family of non-orthogonal time states
\begin{equation}\label{st}
	\ket{t}_C = \sum_{i=0}^{d_C -1 } e^{- i E_i t} \ket{E_i}_C\, , \quad t \in \left[t_0,t_0+T\right) \, ,
\end{equation}
which provide the resolution of the identity
\begin{equation}\label{identityt}
	\mathbb{1}_{C} = \frac{1}{T} \int_{t_0}^{t_0+T} dt\, \ket{t} \bra{t} \, ,
\end{equation}
thus defining a \textit{positive operator-valued measure} (POVM) with elements $\frac{1}{T}\ket{t}\bra{t}dt$.
This framework makes it possible to define time operationally as an internal quantum observable with continuous values, while retaining a discrete and bounded energy spectrum.
Moreover, the rationality condition on energy ratios can be safely relaxed: while the identity (\ref{identityt}) is no longer exact for a generic spectrum, the resulting corrections can be made arbitrarily small, since any real number can be approximated with arbitrary precision by rational ones \cite{nostro3,librotommi,nostro,nostro2,nostro4}.  

The above construction corresponds to a clock with a discrete and bounded spectrum, leading to periodic time states. It can be naturally extended to the case of a continuous and unbounded spectrum for $\hat{H}_C$ \cite{lloydmaccone}. In this case the time states (\ref{st}) can be written:
\begin{equation}
	\ket{t}_C = \int_{-\infty}^{\infty} dE\, e^{-iEt} \ket{E}_C, \quad t \in \left(-\infty,\infty\right) \,,
\end{equation}
and the resolution of the identity becomes $\mathbb{1}_{C} = \frac{1}{2\pi} \int_{-\infty}^{\infty} dt\, \ket{t} \bra{t} $. 
In this case the time states are orthogonal and satisfy $\braket{t|t'}=2\pi \delta(t-t')$.

We emphasize that, throughout the paper, we denote time states as \( \ket{t}_C\) and label time values by \( t \), regardless of whether they are constructed from a discrete or continuous energy spectrum. This will not cause confusion, as the Sections where a continuous and unbounded clock Hamiltonian is used will be clearly specified. 

Turning to $R$ and $S$, we assume the momentum operators $\hat{P}_R$ and $\hat{P}_S$ with bounded, non-degenerate, discrete spectra. The particle $R$ plays the role of a quantum spatial frame, allowing us to define position as a relational observable. In particular, for $R$ we consider $2N_R +1 = d_R$ momentum eigenstates $\ket{p_k}_R$ with eigenvalues
\begin{equation}
	p_k = \frac{2\pi}{L_R} k \,, \quad k = -N_R, -N_R+1, \dots, N_R -1, N_R \,.
\end{equation}
Next, we define the continuous position states
\begin{equation}
	\ket{x}_R = \sum_{k=-N_R}^{N_R} e^{- i p_k x} \ket{p_k}_R, \quad x \in \left[x_0,x_0+L_R\right) \, ,
\end{equation}
satisfying the resolution of the dentity
\begin{equation}
	\mathbb{1}_R = \frac{1}{L_R} \int_{x_0}^{x_0+L_R} dx\, \ket{x} \bra{x}
\end{equation}
and thus representing a POVM for position with infinitesimal non-orthogonal elements $\frac{1}{L_R}\ket{x} \bra{x}dx$.

The same construction applies to the system particle $S$, with $2N_S + 1=d_S$ momentum eigenvalues: 
\begin{equation}
	p_k = \frac{2\pi}{L_S} k, \quad k = -N_S, -N_S+1, \dots, N_S -1, N_S \,.
\end{equation}
The continuous position states are denoted here by:
\begin{equation}
	\ket{y}_S = \sum_{k=-N_S}^{N_S} e^{- i p_k x} \ket{p_k}_S, \quad y \in \left[y_0,y_0+L_S\right)\, .
\end{equation}
Importantly, note that in our framework $\ket{x_0+L_R}_R\equiv\ket{x_0}_R$ and $\ket{y_0+L_S}_S\equiv\ket{y_0}_S$, corresponding to periodic boundary conditions.
In the following, we consistently use \( x \) and \( y \) to denote the position degrees of freedom of the reference \( R \) and the system \( S \), respectively. 

We emphasize that we initially introduced a bounded clock Hamiltonian because our framework is primarily based on the assumption of a finite Universe, with bounded energy for $C$ and compact configuration spaces for both \( R \) and \( S \). The extension to continuous and unbounded clock energy will become useful in Section III.C, where we introduce an interaction potential between \( R \) and \( S \), and in Section VI, where we implement the second quantization formalism. In both cases, we will adopt momentum eigenvalues labeled by all integers $k\in \mathbb{Z}$ (i.e., we will take \( N_R, N_S \to \infty \)), and an unbounded energy spectrum becomes formally necessary. This is reconciled with the physical assumption of a finite Universe by noting that all relevant dynamics are effectively captured within a sufficiently large, though finite, energy window.

	\begin{figure}[t!] 
	\centering 
	\includegraphics [height=6cm]{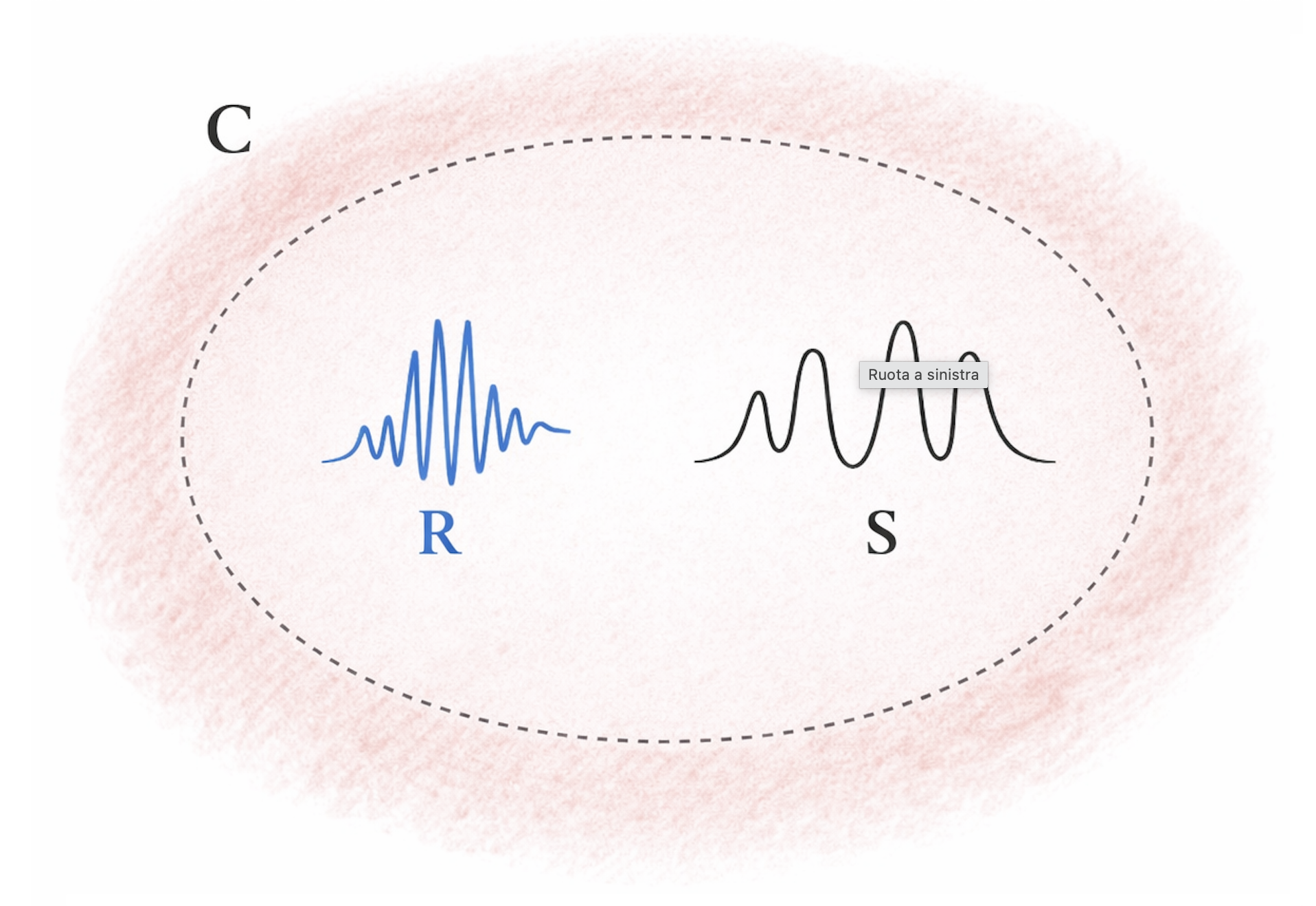} 
	\caption{
		Schematic representation of the relational framework considered in this work. 
		The total system is composed of an environment $C$, which plays the role of a quantum clock, and two subsystems: a reference particle $R$ and a physical system $S$. 
		The dynamics of $S$ is described relationally with respect to $C$ (time) and $R$ (space), without introducing any external classical background. 
		The dashed boundary highlights the subsystem $R+S$, whose conditional state evolves with respect to the clock degrees of freedom encoded in $C$.} 
	\label{image1} 
\end{figure}

\subsection{Spacetime from constraints}
We now present our model of spacetime emerging from entanglement. To this end, we consider a global state $\ket{\Psi} \in \mathcal{H}=\mathcal{H}_C\otimes\mathcal{H}_R\otimes\mathcal{H}_S$ that satisfies the following two constraints simultaneously:
\begin{equation}\label{1}
	\hat{H}\ket{\Psi} = (\hat{H}_C + \hat{H}_R + \hat{H}_S )\ket{\Psi}=0
\end{equation}
and
\begin{equation}\label{2}
	\hat{P}\ket{\Psi} = (\hat{P}_R + \hat{P}_S )\ket{\Psi}=0 \,.
\end{equation}
In equation (\ref{2}) we have assumed $\hat{P}_C = 0$, significantly simplifying the framework. Indeed, while selecting \( \hat{P}_C \ne 0 \) is also possible, it introduces potential limitations on the allowed momenta in order to satisfy both constraints (\ref{1}) and (\ref{2}) simultaneously. Moreover, the clock momentum would explicitly enter the spatial dynamics, preventing the emergence of the standard form of the equations for $S$ alone (for further discussion on this case, we refer the reader to Refs.~\cite{nostro3,librotommi}, where the scenario with \( \hat{P}_C \ne 0 \) is treated). 
In the present analysis, the clock may be identified with an external environment providing a time reference for $R+S$. As discussed previously, this choice is not only physically natural but also particularly effective, since our framework naturally accommodates the use of a generic, non-degenerate Hamiltonian as clock Hamiltonian $\hat{H}_C$. In this setting, the clock carries no spatial momentum, which justifies \( \hat{P}_C = 0 \). 

Starting from the state $\ket{\Psi}$ satisfying (\ref{1}) and (\ref{2}), we expand it over the time basis $\left\{\ket{t}_C\right\}$ using the identity (\ref{identityt}), obtaining
\begin{equation}
	\begin{split}
		\ket{\Psi} &= \frac{1}{T} \int_{t_0}^{t_0 +T} dt \ket{t} \braket{t|\Psi} \\&= \frac{1}{T} \int_{t_0}^{t_0 +T} dt \ket{t}_C \otimes \ket{\phi(t)}_{R,S}
	\end{split}
\end{equation}
where we have defined the \textit{relative state} \cite{everett} (or conditional state) of $R+S$ with respect to $C$ as: $\ket{\phi(t)}_{R,S} = \braket{t|\Psi}$. 
This state satisfies the Schrödinger equation \cite{nostro}:
\begin{equation}\label{evoluzioneRSc}
	i \frac{\partial}{\partial t}\ket{\phi(t)}_{R,S} = \left(\hat{H}_R + \hat{H}_S\right)\ket{\phi(t)}_{R,S}\, ,
\end{equation}
which describes the evolution of the $R+S$ subsystem with respect to the internal clock time $t$.

Similarly, we can expand the state $\ket{\Psi}$ over the position basis $\left\{\ket{x}_R\right\}$ of the reference frame $R$:
\begin{equation}
	\begin{split}
		\ket{\Psi} &= \frac{1}{L_R} \int_{x_0}^{x_0 +L_R} dx \ket{x} \braket{x|\Psi} \\&= \frac{1}{L_R} \int_{x_0}^{x_0 +L_R} dx \ket{x}_R \otimes \ket{\varphi(x)}_{C,S}
	\end{split}
\end{equation}
where the conditional state of $C+S$ is $\ket{\varphi(x)}_{C,S} = \braket{x|\Psi}$.
For such state it is possible to show that \cite{nostro3,librotommi}:
\begin{equation}\label{fondamentale1}
	\hat{P}_S \ket{\varphi(x)}_{C,S} = i \frac{\partial}{\partial x} \ket{\varphi(x)}_{C,S}\,, 
\end{equation}
confirming that $\hat{P}_S$ acts as the generator of translations in the spatial coordinate $x$ for $\ket{\varphi(x)}_{C,S}$. 

We can now expand $\ket{\Psi}$ simultaneously in the bases $\left\{\ket{t}_C\right\}$ and $\left\{\ket{x}_R\right\}$, obtaining:
\begin{equation}\label{Psiespanso}
	\begin{split}
		\ket{\Psi} &= \left( \frac{1}{T}\int_{t_0}^{t_0 + T} dt \ket{t}\bra{t} \otimes \frac{1}{L_R}\int_{x_0}^{x_0 + L_R} dx \ket{x}\bra{x} \right) \ket{\Psi} 
		\\& = \frac{1}{T} \frac{1}{L_R} \int_{t_0}^{t_0 + T} dt \int_{x_0}^{x_0 + L_R} dx \ket{t}_C \otimes \ket{x}_R \otimes \ket{\psi(x,t)}_S
	\end{split}
\end{equation}
where we define $\ket{\psi(x,t)}_S = \braket{t,x|\Psi}$ as the relative state of the system $S$ at clock time $t$ and conditioned on the position $x$ of the reference frame $R$. 

The conditional probability density of measuring the position $y$ of $S$ at time $t$, given that $R$ is in position $x$, is
\begin{multline}\label{probfinale}
	P(y \: \text{on} \: S\:|\:x \: \text{on} \: R,\: t \: \text{on} \: C) \\ \equiv P(y-x,t)  \propto \left| \braket{y|\psi(x,t)} \right|^2 \,,
\end{multline}
where the proportionality sign is left, as the overall prefactor is fixed by the chosen normalization conventions.
Through entanglement, we thus obtain a conditional probability density for $S$ that encodes its evolution in both time and space. Time is defined via the internal clock $C$, while space is characterized by the relative distance between $S$ and $R$. Indeed, we emphasize that the momentum constraint (\ref{2}) ensures that the probability distribution (\ref{probfinale}) and, more generally, the relative state (\ref{Psiespanso}) depend only on the spatial separation $y - x$ between the system and the reference frame. This key feature will be crucial for the derivation of the equations in what follows. For probabilities involving multiple time measurements, we again directly refer to Refs.~\cite{nostro3,librotommi}.

Finally, we notice that the property (\ref{fondamentale1}) holds also for the state $\ket{\psi(t,x)}_S = \braket{t,x|\Psi}=\braket{x|\phi(t)}_{R,S}$, namely: 	
\begin{equation}\label{fondamentale2}
	\hat{P}_S \ket{\psi(x,t)}_{S} = i \frac{\partial}{\partial x} \ket{\psi(x,t)}_{S}\,.
\end{equation}
Furthermore, due to the underlying symmetry of the formalism between \( R \) and \( S \), the same reasoning applies if we exchange their roles, and in particular the following property also holds:
\begin{equation}\label{fondamentale3}
	\hat{P}_R \ket{\psi(y,t)}_{R} = i \frac{\partial}{\partial y} \ket{\psi(y,t)}_{R}
\end{equation}
where $\ket{\psi(y,t)}_{R} =\braket{t,y|\Psi}= \braket{y|\phi(t)}_{R,S}$.
Also relations (\ref{fondamentale2}) and (\ref{fondamentale3}) will play a key role and will be used multiple times in the following Sections.

\section{Schrödinger wave equation}
In this Section, we derive and analyze the emergence of the Schrödinger wave equation within our relational framework. We begin by considering the case in which the kinetic energy of the reference particle \( R \) can be neglected, allowing us to study the effective dynamics of the system \( S \) alone. We then move on to the more general case in which the kinetic contribution of \( R \) is explicitly taken into account and, finally, we investigate how an interaction potential depending on the distance between \( S \) and \( R \) can be incorporated into the formalism.

\subsection{Schrödinger equation for $S$}
We begin by considering \( R \) and \( S \) as non-relativistic free particles, with energies \( \hat{H}_R = \hat{P}^2_R/2M \) and \( \hat{H}_S = \hat{P}^2_S/2m \). Under the assumption \( M \gg m, |p_k| \,\, \forall k \), the energy constraint (\ref{1}) reduces to the following form:
\begin{equation}\label{energyS}
	\left(\hat{H}_C + \frac{\hat{P}^2_S}{2m}\right)\ket{\Psi} \approx 0 \, .
\end{equation}
Projecting this equation onto the basis \( \{\ket{t}_C\} \) in the clock subspace yields the evolution equation for the \( R+S \) state:
\begin{equation}\label{27b}
	i \frac{\partial}{\partial t}\ket{\phi(t)}_{R,S} \approx \frac{\hat{P}^2_S}{2m} \ket{\phi(t)}_{R,S} \, .
\end{equation}
We then evaluate (\ref{27b}) in the position basis \( \{\ket{x}_R\} \) of the reference subspace. Using relation (\ref{fondamentale2}), we obtain the equation for the relative state of \( S \):
\begin{equation}\label{Sket}
	i \frac{\partial}{\partial t}\ket{\psi(x,t)}_S \approx -\frac{1}{2m} \frac{\partial^2}{\partial x^2} \ket{\psi(x,t)}_S \,.
\end{equation}
We note that equation (\ref{Sket}) has the form of the Schrödinger equation for a free particle of mass \( m \), but it governs the evolution of a ket in the \( S \) subspace, with \( x \) and \( t \) entering as parameters due to the imposed constraints. To obtain the Schrödinger equation, we project (\ref{Sket}) onto the basis \( \{\ket{y}_S\} \), yielding:
\begin{equation}\label{27}
	i \frac{\partial}{\partial t} \psi(y - x, t) \approx  -\frac{1}{2m} \frac{\partial^2}{\partial x^2} \psi(y - x, t)\, ,
\end{equation}
where the dependence of the wave function on \( y - x \) is ensured by the constraint (\ref{2}). As a final step, we introduce the new spatial coordinate \( \xi = y - x \), which naturally reflects the relational character of the position of \( S \) with respect to the reference particle \( R \). The derivatives with respect to \( x \) transform as follows:
\begin{equation}\label{derivatex}
	\left\{
	\begin{aligned}
		& \frac{\partial}{\partial x} = \frac{\partial \xi}{\partial x} \frac{\partial}{\partial \xi} = - \frac{\partial}{\partial \xi} \\
		& \frac{\partial^2}{\partial x^2} = \frac{\partial}{\partial x} \left( \frac{\partial \xi}{\partial x} \frac{\partial}{\partial \xi} \right) =  \frac{\partial^2}{\partial \xi^2}
	\end{aligned}
	\right.
\end{equation}
so that equation (\ref{27}) becomes
\begin{equation}\label{eqS}
	i \frac{\partial}{\partial t} \psi(\xi, t) \approx  -\frac{1}{2m} \frac{\partial^2}{\partial \xi^2} \psi(\xi, t)\,.
\end{equation}
In equation (\ref{eqS}), the wave function is expressed in terms of the relational variable \( \xi \), which directly incorporates the relational aspect of the dynamics of $S$.

We do not explicitly solve equation (\ref{eqS}), as its solution clearly corresponds to the well-known wave function of a free Schrödinger particle. Rather, our aim is to recover such solution directly from the underlying global constraints of the framework, showing how the familiar dynamics emerges from purely relational principles.

Before proceeding further, we make some preliminary considerations regarding the construction of the global state \( \ket{\Psi} \). In order to satisfy both the momentum and energy constraints, the state must be built so that no eigenvalue of the system \( S \) is left uncorrelated with the corresponding degrees of freedom of \( R \) and \( C \). For the momentum constraint to be satisfied, it is necessary to ensure that each momentum eigenvalue of \( S \) can be paired with a corresponding (opposite) momentum eigenvalue of \( R \). This can be achieved under the assumption of equal spectra for \( \hat{P}_R \) and \( \hat{P}_S \) (i.e., $N_R=N_S$, $L_R=L_S$), or (more generally) by requiring \( d_R \gg d_S \) and \( L_R \gg L_S \), so that the spectrum of \( \hat{P}_R \) is sufficiently broad and dense to match all values of the momentum \( \hat{P}_S \). Similarly, ensuring that no energy eigenstate of \( R+S \) (or of \( S \) alone, if the energy of \( R \) is neglected in the constraint) is excluded from the dynamics requires the clock \( C \) to have a sufficiently large and finely spaced energy spectrum, as discussed in Refs.~\cite{nostro,nostro2,librotommi}, where the conditions for a \lq\lq good clock\rq\rq\:are presented and analyzed in detail.

It is worth noting that all the requirements mentioned are automatically fulfilled in the case of continuous and unbounded energy spectrum for $C$ and continuous unbounded momentum spectra for $R$ and $S$. 
Nonetheless, we never make use of continuous and unbounded momentum spectra throughout this work.

Assuming, then, that all the conditions discussed above are satisfied, the general state \( \ket{\Psi} \) constrained by (\ref{2}) and (\ref{energyS}) can be written:
\begin{equation}\label{32}
	\ket{\Psi} = \sum_{k=-N_S}^{N_S} c_k \ket{E = -\frac{p^2_k}{2m}}_C \otimes \ket{p = -p_k}_R \otimes \ket{p_k}_S \,.
\end{equation}
By projecting (\ref{32}) onto the time basis $\{\ket{t}_C\}$ and the position basis $\{\ket{x}_R\}$ in $C$ and $R$ respectively, we obtain the expression for the relative state of $S$ $\ket{\psi(x,t)}_S = \braket{t,x|\Psi}$, namely:
\begin{equation}
	\ket{\psi(x,t)}_S = \sum_{k=-N_S}^{N_S} c_k e^{-i \frac{p^2_k}{2m} t} e^{-i p_k x} \ket{p_k}_S \, ,
\end{equation}
leading to 
\begin{equation}
	\begin{split}
		\psi(y-x,t) &= \braket{y|\psi(x,t)}_S \\&
		= \sum_{k=-N_S}^{N_S} c_k e^{-i \frac{p^2_k}{2m} t} e^{i p_k (y-x)}
	\end{split}
\end{equation}
and finally to
\begin{equation}
	\psi(\xi,t) = \sum_{k=-N_S}^{N_S} c_k e^{i ( p_k\xi - \frac{p^2_k}{2m} t )}
\end{equation}
where we have introduced again the coordinate $\xi=y-x$.

To further constrain the form of the coefficients \( c_k \), we now impose the normalization condition of the wave function. For simplicity—both here and in the remainder of the paper—we set $N_R=N_S$, \( x_0 = y_0 = 0 \) and \( L_R = L_S \equiv L \), implying equal spectra for \( \hat{P}_R \) and \( \hat{P}_S \). 

We reiterate that, in our framework, both the reference \( R \) and the system \( S \) are defined on a compact configuration space of length \( L \), with positions \( x,y \in [0, L) \) and periodic boundary conditions, implying a cyclical repetition of the states \( \ket{x}_R \) and \( \ket{y}_S \) under translations by integer multiples of \( L \).
Accordingly, any wave function depending on the relational coordinate \( \xi = y - x \) also repeats cyclically under translations by integer multiples of \( L \), so that the physics is fully captured within any such interval. We thus choose the domain $\xi \in \left[-L/2, L/2\right)$ as a natural choice centered around the origin $\xi=0$ ($y=x$). 
Specifically, we ask here:

\begin{equation}\label{36}
	\begin{split}
		&\int_{-L/2}^{L/2} d\xi\: \left| \psi (\xi,t)\right|^2 = \\& 
		= 	\int_{-L/2}^{L/2} d\xi\: \sum_{k=-N_S}^{N_S} \sum_{n=-N_S}^{N_S} c_k c^{*}_n e^{i ( p_k\xi - \frac{p^2_k}{2m} t )} e^{-i ( p_n\xi - \frac{p^2_n}{2m} t )} \\&
		= \sum_{k=-N_S}^{N_S} \sum_{n=-N_S}^{N_S} c_k c^{*}_n e^{-i(\frac{p^2_k}{2m} - \frac{p^2_n}{2m})t} 	\int_{-L/2}^{L/2} d\xi\: e^{i\xi(p_k-p_n)}\\&
		= L  \sum_{k=-N_S}^{N_S} |c_k|^2 =1
	\end{split}
\end{equation}
where we have used 
\begin{equation}
	\int_{-L/2}^{L/2} d\xi\: e^{i\xi(p_k-p_n)} = L\delta_{k,n} \,.
\end{equation}
Equation (\ref{36}) leads to $c_k = \frac{1}{\sqrt{L}} a_k$, which allow us to finally obtain:
\begin{equation}
	\psi(\xi,t) = \sum_{k=-N_S}^{N_S} a_k u_k(\xi,t)
\end{equation}
with
\begin{equation}
	u_k(\xi,t) = \frac{1}{\sqrt{L}}e^{i(p_k\xi - \frac{p^2_k}{2m}t)} \, ,
\end{equation}
which is the correct solution for the wave function of a free Schrödinger particle and exactly satisfies (\ref{eqS}).

\subsection{Schrödinger equation for $R+S$}
We now consider the case in which the energy of \( R \) cannot be neglected. In this setting, the energy constraint (\ref{1}) takes the form:
\begin{equation}\label{enRS}
	\left(\hat{H}_C + \frac{\hat{P}^2_R}{2M} + \frac{\hat{P}^2_S}{2m}\right)\ket{\Psi}=0 \,.
\end{equation}
Projecting onto the basis \( \{\ket{t}_C\} \) in the clock subspace yields the evolution equation for the joint state of \( R +S \):
\begin{equation}\label{40b}
	i\frac{\partial}{\partial t} \ket{\phi(t)}_{R,S} = 	\left(\frac{\hat{P}^2_R}{2M} + \frac{\hat{P}^2_S}{2m}\right) \ket{\phi(t)}_{R,S} \,.
\end{equation}
We then evaluate (\ref{40b}) the position bases \( \{\ket{x}_R\} \) and \( \{\ket{y}_S\} \), and apply relations (\ref{fondamentale2}) and (\ref{fondamentale3}), obtaining:
\begin{equation}\label{40}
	i\frac{\partial}{\partial t} \psi(y-x,t) = \left( -\frac{1}{2M} \frac{\partial^2}{\partial y^2} - \frac{1}{2m} \frac{\partial^2}{\partial x^2} \right) \psi(y-x,t) \,.
\end{equation}
We now introduce again the coordinate \( \xi = y - x \). Along with the derivative relations in (\ref{derivatex}), we also need:
\begin{equation}\label{derivatey}
	\left\{
	\begin{aligned}
		& \frac{\partial}{\partial y} = \frac{\partial \xi}{\partial y} \frac{\partial}{\partial \xi} = \frac{\partial}{\partial \xi} \\
		& \frac{\partial^2}{\partial y^2} = \frac{\partial}{\partial y} \left( \frac{\partial \xi}{\partial y} \frac{\partial}{\partial \xi} \right) =  \frac{\partial^2}{\partial \xi^2} .
	\end{aligned}
	\right.
\end{equation}
By inserting the relations (\ref{derivatex}) and (\ref{derivatey}) into equation (\ref{40}), we finally obtain:
\begin{equation}\label{finaleRS}
		i\frac{\partial}{\partial t}	\psi(\xi,t) 
		=  - \frac{1}{2\mu} \frac{\partial^2}{\partial \xi^2} \psi(\xi,t) \, ,
\end{equation}
where we have introduced the reduced mass $\mu = \frac{mM}{m+M}$.
This equation shows that the composite system \( R+S \), when described in terms of \( \xi = y - x \), behaves effectively as a single free particle with reduced mass $\mu$, evolving in time according to the standard Schrödinger dynamics—thus highlighting how relational degrees of freedom can encapsulate the full dynamics of the joint system.

As in previous paragraph, we do not explicitly solve equation (\ref{finaleRS}) but we recover the solution directly from constraints (\ref{2}) and (\ref{enRS}).
The general form of the global state satisfying both constraints can be written:
\begin{equation}
	\ket{\Psi} = \sum_{k=-N_S}^{N_S} c_k \ket{E = - \frac{p^2_k}{2\mu}}_C \otimes \ket{p = -p_k}_R \otimes \ket{p_k}_S \,.
\end{equation}
Therefore, by following the same steps as in the previous paragraph (projecting the global state onto the bases \( \{\ket{t}_C\} \), \( \{\ket{x}_R\} \), \( \{\ket{y}_S\} \), introducing the coordinate \( \xi \), and imposing \( \int_{-L/2}^{L/2} d\xi\, |\psi(\xi,t)|^2 = 1 \)), we obtain again: $	\psi(\xi,t) = \sum_{k=-N_S}^{N_S} a_k u_k(\xi,t)$ with
\begin{equation}
	u_k(\xi,t) = \frac{1}{\sqrt{L}}e^{i(p_k\xi - \frac{p^2_k}{2\mu}t)} \,. 
\end{equation}
This solution corresponds to the standard wave function of a free particle with mass \( \mu \), thus confirming the expected form for the joint dynamics of the \( R+S \) system.

\subsection{Introducing the potential}
In this paragraph, we show how the formalism naturally allows for the introduction of an interaction potential between the reference \( R \) and the system \( S \). Throughout this analysis, we assume $R$ and $S$ with unbounded momentum spectra, namely we take $N_R,N_S \to \infty$. Within this framework, we can introduce the Hermitian operators \( \hat{X}=\frac{1}{L}\int_{-L/2}^{L/2}dx\:x\ket{x}\bra{x} \) and \( \hat{Y}=\frac{1}{L}\int_{-L/2}^{L/2}dy\:y\ket{y}\bra{y} \), describing the positions of \( R \) and \( S \), and consequently we can define an interaction potential \( V( \hat{Y} - \hat{X}) \). Note that, since the momenta $p_k$ in $R$ and $S$ span all integer multiples of \( 2\pi/L \), we also assume here a continuous and unbounded energy spectrum for $C$, so that the global energy constraint can be exactly satisfied.

Such constraint reads:
\begin{equation}
	\left(\hat{H}_C + \frac{\hat{P}^2_R}{2M} + \frac{\hat{P}^2_S}{2m} + V(\hat{Y}-\hat{X})\right)\ket{\Psi}=0 
\end{equation}
which, upon projection onto the bases \( \{\ket{t}_C\} \), \( \{\ket{x}_R\} \), \( \{\ket{y}_S\} \), and applying relations (\ref{fondamentale2}) and (\ref{fondamentale3}), gives:
\begin{multline}
	i\frac{\partial}{\partial t} \psi(y-x,t) = \\ = \left( -\frac{1}{2M} \frac{\partial^2}{\partial y^2} - \frac{1}{2m} \frac{\partial^2}{\partial x^2} + V(y-x) \right) \psi(y-x,t) \,.
\end{multline}
Introducing then the relational coordinate \( \xi \), and using the transformations for the derivatives, the equation takes the final form:
\begin{equation}\label{finaleV}
	i\frac{\partial}{\partial t}	\psi(\xi,t) = - \frac{1}{2\mu} \frac{\partial^2}{\partial \xi^2} \psi(\xi,t) + V(\xi)\psi(\xi,t) \, ,
\end{equation}
where we have introduced again $\mu = \frac{mM}{m+M}$. We do not attempt to solve equation (\ref{finaleV}), as its solution depends on the specific form of the chosen potential \( V(\xi) \). Our goal here was simply to demonstrate that an interaction potential between \( R \) and \( S \) can be naturally incorporated within the formalism. It is also worth noting that, for \( M \gg m \), we have \( \mu \approx m \), and the equation reduces to that of a single particle \( S \) moving in a potential that depends on its position relative to the referencce \( R \).

\section{Klein-Gordon wave equation}
In this Section, we derive and analyze the Klein-Gordon wave equation within our relational framework. 
We focus on the case where the contribution of the reference particle \( R \) to the total energy can be neglected, in order to avoid additional technical complications, which go beyond the scope of the present work. We thus obtain the effective equation governing the dynamics of the system \( S \) alone.

Among the cases considered in this work, the Klein–Gordon equation presents additional challenges due to the second-order nature of its time derivative, which complicates both its interpretation and its formulation within the relational framework. In this context, one could consider constraints involving the squares of the energy operators. However, in the present work we instead adopt a constraint linear in the energy operators, as this choice is more directly aligned with the PaW construction, where the global constraint arises as the sum of the subsystem energies, placing the total state in an eigenstate of the total energy. Within this setting, the Klein–Gordon structure is recovered by introducing two independent constraints accounting for the positive- and negative-energy sectors. In light of these considerations, we note that we are deviating—both conceptually and formally—from the Klein-Gordon treatment briefly outlined in Refs.~\cite{nostro3,librotommi}, where the reference $R$ is assumend to be non-relativistic.

We thus consider (c=1): $\hat{H}_R = \sqrt{\hat{P}^2_R + M^2}$ and $\hat{H}_S = \sqrt{\hat{P}^2_S + m^2}$. For $M \gg m, |p_k|\,\, \forall k$, we can write: 
\begin{equation}
	\hat{H}_R \simeq M + \frac{\hat{P}^2_R}{2M} \approx M \,.
\end{equation}
Under this assumption, the reference \( R \) enters the energy constraint (to be introduced shortly) with a constant mass term \( M \), which can be neglected as it contributes only an unobservable global phase factor to the dynamics of the relative state of the system \( S \). Namely, we can consider the global state satisfying (\ref{2}) and:
\begin{equation}\label{enKG}
	\left( \hat{H}_C \pm  \sqrt{\hat{P}^2_S + m^2} \right)\ket{\Psi_{\pm}} \approx 0  \, ,
\end{equation}
where we are treating together both energy constraints since they lead to the same final equation for the particle $S$. Projecting the equation  first onto the basis \( \{\ket{t}_C\} \) and then onto the basis \( \{\ket{x}_R\} \), we obtain:
\begin{equation}
	i\frac{\partial}{\partial t} \ket{\phi_{\pm}(t)}_{R,S} \approx \pm \sqrt{\hat{P}^2_S + m^2} \ket{\phi_{\pm}(t)}_{R,S}
\end{equation}
and
\begin{equation}
	i\frac{\partial}{\partial t} \ket{\psi_{\pm}(x,t)}_S \approx \pm \sqrt{\hat{P}^2_S + m^2} \ket{\psi_{\pm}(x,t)}_S \,.
\end{equation}
We now iterate the evolution equation for the relative state of \( S \) in order to obtain
\begin{equation}
	- \frac{\partial^2}{\partial t^2} \ket{\psi_{\pm}(x,t)}_S \approx \left( \hat{P}^2_S + m^2 \right) \ket{\psi_{\pm}(x,t)}_S 
\end{equation}
which, through (\ref{fondamentale2}), can be cast in the form:
\begin{equation}
	\left(  \frac{\partial^2}{\partial t^2} - \frac{\partial^2}{\partial x^2} +m^2 \right)\ket{\psi_{\pm}(x,t)}_S  \approx 0 \,.
\end{equation}
By further projecting onto the position basis \( \{\ket{y}_S\} \) and introducing the relative coordinate \( \xi = y - x \), we finally obtain:
\begin{equation}
	\left(  \frac{\partial^2}{\partial t^2} - \frac{\partial^2}{\partial x^2} + m^2 \right)\psi_{\pm}(y-x,t) \approx 0
\end{equation}
and
\begin{equation}
	\left(  \frac{\partial^2}{\partial t^2} - \frac{\partial^2}{\partial \xi^2} + m^2 \right)\psi_{\pm}(\xi,t) \approx 0 \, ,
\end{equation}
which is the Klein-Gordon equation expressed in terms of the relative spatial coordinate between \( R \) and \( S \). As in the previous Section, we will explicitly derive its solution directly from the underlying constraint structure.

The global state, satisfying both (\ref{2}) and (\ref{enKG}), can be written:
\vspace{-0.2cm}
\begin{equation}
	\ket{\Psi_{\pm}} = \sum_{k=-N_S}^{N_S} c_{\pm,k} \ket{E=\mp \epsilon_k}_C\otimes \ket{p=-p_k}_R\otimes\ket{p_k}_S  \, ,
\end{equation}
where we have introduced $\epsilon_k=\sqrt{p^2_k + m^2}$.
By expressing $\ket{\Psi_{\pm}} $ in the time basis \( \{\ket{t}_C\} \) and position basis \( \{\ket{x}_R\} \) of subsystems \( C \) and \( R \), we obtain the expression for the relative state $\ket{\psi_{\pm}(x,t)}_S = \braket{t,x|\Psi_{\pm}}$ of the system $S$, namely:
\begin{equation}
	\ket{\psi_{\pm}(x,t)}_S = \sum_{k=-N_S}^{N_S} c_{\pm,k} e^{\mp i \epsilon_k t} e^{-i p_k x} \ket{p_k}_S \, ,
\end{equation}
leading to $\psi_{\pm}(y-x,t) = \braket{y|\psi_{\pm}(x,t)}_S = \sum_{k=-N_S}^{N_S} c_{\pm,k} e^{\mp i \epsilon_k t} e^{i p_k (y-x)}$
and finally to
\begin{equation}
	\psi_{\pm}(\xi,t) = \sum_{k=-N_S}^{N_S} c_{\pm,k} e^{i ( p_k\xi \mp \epsilon_k t )} \, ,
\end{equation}
where we have reintroduced the coordinate $\xi=y-x$.

In order to gain additional information about the structure of the coefficients \( c_{\pm,k} \), we now employ the expression for the temporal component of the conserved Klein-Gordon four-current (see for example \cite{peskin}):
\begin{equation}
	\rho_{\pm} =i \left(  \psi^{*}_{\pm}(\xi,t) \frac{\partial}{\partial t} \psi_{\pm}(\xi,t) - \psi_{\pm}(\xi,t) \frac{\partial}{\partial t} \psi^{*}_{\pm}(\xi,t)\right)
\end{equation}
and we require $\int_{-L/2}^{L/2} d\xi\:\rho_{\pm}=\pm 1$. 
We make the calculation for the positive-energy solution, namely we have:
\begin{widetext}
	\begin{equation}
		\begin{split}
			\rho_{+} &= \sum_{k=-N_S}^{N_S}\sum_{n=-N_S}^{N_S} c^{*}_{+,k} c_{+,n} \epsilon_n e^{i\xi(p_n-p_k)}e^{-it(\epsilon_n-\epsilon_k)}  +   \sum_{k=-N_S}^{N_S}\sum_{n=-N_S}^{N_S} c^{*}_{+,k} c_{+,n} \epsilon_k e^{i\xi(p_n-p_k)}e^{-it(\epsilon_n-\epsilon_k)} \\ \\&
			=   \sum_{k=-N_S}^{N_S}\sum_{n=-N_S}^{N_S} c^{*}_{+,k} c_{+,n} (\epsilon_n + \epsilon_k)  e^{i\xi(p_n-p_k)}e^{-it(\epsilon_n - \epsilon_k)} 
		\end{split}
	\end{equation}
	leading to
	\begin{equation}\label{64}
		\int_{-L/2}^{L/2} d\xi\: \rho_+ =  \int_{-L/2}^{L/2} d\xi\: \left(  \sum_{k=-N_S}^{N_S}\sum_{n=-N_S}^{N_S} c^{*}_{+,k} c_{+,n} (\epsilon_n + \epsilon_k)  e^{i\xi(p_n-p_k)}e^{-it(\epsilon_n - \epsilon_k)}\right)
		= L \sum_{k=-N_S}^{N_S} |c_{+,k}|^2 2 \epsilon_k = 1 \,.
	\end{equation}
\end{widetext}
We can thus define the coefficients $a_k$ through: $c_{+,k} = \frac{1}{\sqrt{L}\sqrt{2\epsilon_k}} a_k $,
with which we can write
\begin{equation}\label{kgpositive}
	\psi_{+}(\xi,t) =  \sum_{k=-N_S}^{N_S}  a_k u_{+,k}(\xi,t) \, ,
\end{equation}
where
\begin{equation}\label{61}
	u_{+,k}(\xi,t) =  \frac{1}{\sqrt{L}\sqrt{2\epsilon_k}} e^{i(p_k\xi - \epsilon_k t)}\,. 
\end{equation}
Analogously, we find:
\begin{equation}\label{bstar}
	\psi_{-}(\xi,t) =  \sum_{k=-N_S}^{N_S}  b^{*}_k u_{-,k}(\xi,t)
\end{equation}
where
\begin{equation}\label{63}
	u_{-,k}(\xi,t) = \frac{1}{\sqrt{L}\sqrt{2\epsilon_k}} e^{i(p_k\xi + \epsilon_k t)}\,. 
\end{equation}
These results correspond to the well-known solutions of the Klein-Gordon equation, in agreement with the standard treatment presented (for example) in Ref.~\cite{greinerrelativistic}.

We note that, in equation (\ref{bstar}), we wrote the negative-energy solution introducing the coefficients $b^{*}_k$ to set the stage for their promotion to creation operators in the second quantized theory. In this sense, we can also make a further step by assuming $N_R,N_S\to \infty$ in $R$, $S$ and $\hat{H}_C$ with continuous unbounded spectrum. We thus define 
$u_k (\xi,t) \equiv u_{+,k}(\xi,t)$ 
and rewrite (\ref{kgpositive}) and (\ref{bstar}) as:
\begin{equation}
	\psi_{+}(\xi,t) =  \sum_{k=-\infty}^{\infty}  a_k u_{k}(\xi,t)
\end{equation}
and
\begin{equation}\label{kgnegative2}
	\psi_{-}(\xi,t) =  \sum_{k=-\infty}^{\infty}  b^{*}_k u^{*}_{k}(\xi,t) \, ,
\end{equation}
where $u^{*}_{k}(\xi,t)= u_{-,-k}(\xi,t)$, namely 
\begin{equation}
	u^{*}_{k}(\xi,t) = \frac{1}{\sqrt{L}\sqrt{2\epsilon_k}} e^{- i(p_k\xi - \epsilon_k t)}\, . 
\end{equation}
In writing (\ref{kgnegative2}) we used \(\epsilon_k = \epsilon_{-k}\). We will adopt this formalism in Section~VI, where we quantize the field.

\section{Dirac wave equation}
In this Section, we derive and study the Dirac wave equation assuming the system \( S \) as spin-\(1/2\) particle. We adopt a similar approximation as in the Klein-Gordon case, neglecting the energy contribution of the reference system \( R \) (assumed to be, as in the previous Section, a spin-\(0\)~particle with $\hat{H}_R=\sqrt{\hat{P}^2_R + M^2}$ and $M>>m,|p_k|\,\, \forall k$) in the constraint. This allows us again to isolate and derive the effective relativistic dynamics for \( S \) and, in contrast to the Klein-Gordon case, we will see that a single energy constraint is sufficient to obtain both the positive- and negative-energy solutions.

In $1+1$ spacetime, the Hamiltonian $\hat{H}_S$ can be written: 
\begin{equation}
	\hat{H}_S = \hat{P}_S \sigma_1 + m\sigma_3 \,,
\end{equation}
where we denote as $\sigma_1$ and $\sigma_3$ the Pauli matrices $\sigma_x$ and $\sigma_z$, respectively. Accordingly, the energy constraint reads:
\begin{equation}\label{enDirac}
	\left( \hat{H}_C + \hat{P}_S \sigma_1 + m\sigma_3 \right)\ket{\Psi} \approx 0 
\end{equation}
which, when represented in the \( \{ \ket{t}_C \} \) basis, leads to 
\begin{equation}\label{72}
	i\frac{\partial}{\partial t} \ket{\phi(t)}_{R,S} \approx \left( \hat{P}_S \sigma_1 + m\sigma_3\right) \ket{\phi(t)}_{R,S} \,.
\end{equation}
Evaluating the equation in the basis \( \{ \ket{x}_R \} \) and applying relation (\ref{fondamentale2}), we obtain:
\begin{equation}
	i\frac{\partial}{\partial t} \ket{\psi(x,t)}_S \approx  \left(	i\frac{\partial}{\partial x} \sigma_1 + m\sigma_3 \right) \ket{\psi(x,t)}_S \,.
\end{equation} \\[1em]
Projecting finally onto the position basis \( \{\ket{y}_S\} \) and introducing the coordinate \( \xi = y - x \), we arrive at:
\begin{equation}\label{eqDirac}
	i\frac{\partial}{\partial t} \ket{\psi(\xi,t)}_{\sigma} \approx - i\frac{\partial}{\partial \xi} \sigma_1 \ket{\psi(\xi,t)}_{\sigma} + m\sigma_3 \ket{\psi(\xi,t)}_{\sigma} 
\end{equation}
which corresponds to the Dirac equation for the system $S$ expressed in terms of the relative coordinate \( \xi \). Note that this equation is still written in ket form, as it retains the two-component spinor structure (labeled by $\sigma$), which has not yet been resolved by choosing a representation.
The state $\ket{\psi(\xi,t)}_{\sigma}$ is therefore a two-component object, which can be expressed more conveniently as:
\begin{equation}\label{statovett}
	\ket{\psi(\xi,t)}_{\sigma}\equiv \psi(\xi,t) = \begin{pmatrix}
		\psi_{1}(\xi,t) \\
		\psi_{2}(\xi,t) 
	\end{pmatrix} \,.
\end{equation}
Using this notation, equation (\ref{eqDirac}) can be rewritten:
\begin{equation}\label{eqDirac2}
	\left( i \gamma^0 \frac{\partial}{\partial t} + i \gamma^1 \frac{\partial}{\partial \xi} - m \right) 	\psi(\xi,t) \approx 0 \, ,
\end{equation}
where we have introduced the matrices $\gamma^0 = \sigma_3$ and $\gamma^1=i\sigma_2$, with $\sigma_2$ denoting the Pauli matrix $\sigma_y$ (notice that $\gamma^0$, $\gamma^1$ satisfy $\left(\gamma^0\right)^2=\mathbb{1}$, $\left(\gamma^1\right)^2=- \mathbb{1}$, $\{\gamma^0,\gamma^1\}=0$).

We can now derive the solution of equations (\ref{eqDirac}) and (\ref{eqDirac2}) starting from the global constraints of our model. The global state satisfying the momentum constraint (\ref{2}) can be written:
\begin{equation}
	\ket{\Psi} = \sum_{\sigma=1}^{2} \sum_{n=0}^{d_C - 1}\sum_{k=-N_S}^{N_S} c^{(\sigma)}_{n,k} \ket{E_n}_C\otimes\ket{p=-p_k}_R\otimes\ket{p_k,\sigma}_S \, ,
\end{equation}
where the spinor degree of freedom is made explicit~\cite{dirac,librodirac} in the ket of the system \( S \). We now determine the allowed energy values for $S$ by imposing (\ref{enDirac}), namely:
\begin{widetext}
	\begin{multline}\label{76}
		\left( \hat{H}_C + \hat{P}_S \sigma_1 + m\sigma_3 \right) \ket{\Psi} 
		= \sum_{n=0}^{d_C-1}\sum_{k=-N_S}^{N_S} \left(c^{(1)}_{n,k}E_n + c^{(2)}_{n,k}p_k + c^{(1)}_{n,k}m \right) \ket{E_n}_C\otimes\ket{p=-p_k}_R\otimes\ket{p_k,1}_S +\\+
		\sum_{n=0}^{d_C-1}\sum_{k=-N_S}^{N_S} \left(c^{(2)}_{n,k}E_n + c^{(1)}_{n,k}p_k - c^{(2)}_{n,k}m \right) \ket{E_n}_C\otimes\ket{p=-p_k}_R\otimes\ket{p_k,2}_S \approx 0 \,.
	\end{multline}
\end{widetext}
To solve equation (\ref{76}), we impose that both terms vanish, which yields the condition:
\begin{equation}
	\left\{
	\begin{aligned}
		& c^{(1)}_{n,k}(E_n + m) + c^{(2)}_{n,k}p_k= 0\\
		& c^{(1)}_{n,k}p_k + c^{(2)}_{n,k}(E_n -m)= 0 \,.
	\end{aligned}
	\right.
\end{equation}
Such system has non-trivial solutions only in the case of vanishing determinant of the coefficients: 
\begin{equation}
	\textit{det} \begin{pmatrix}
		(E_n + m) & p_k \\ p_k & (E_n -m)
	\end{pmatrix} = 0
\end{equation}
leading to $\left( E_n + m\right)\left( E_n - m\right) - p^2_k = 0$
and finally to 
\begin{equation}
	E_n = \pm \sqrt{p^2_k + m^2} \, .
\end{equation}
The global state $\ket{\Psi}$ contains now two energy solutions and can thus be rewritten:
\begin{equation}\label{78}
	\ket{\Psi_{\pm}} = \sum_{\sigma=1}^{2} \sum_{k=-N_S}^{N_S} c^{(\sigma)}_{\pm,k} \ket{E=\mp \epsilon_k}_C\otimes\ket{p=-p_k}_R\otimes\ket{p_k,\sigma}_S
\end{equation}
where we have introduced again $\epsilon_k=\sqrt{p^2_k + m^2}$. 

We project equation (\ref{78}) in the bases \( \{\ket{t}_C\} \) and \( \{\ket{x}_R\} \), in order to obtain the expression for the relative state $\ket{\psi_{\pm}(x,t)}_S = \braket{t,x|\Psi_{\pm}}$. We have:
\begin{equation}
	\ket{\psi_{\pm}(x,t)}_S = \sum_{\sigma=1}^{2} \sum_{k=-N_S}^{N_S} c^{(\sigma)}_{\pm ,k} e^{\mp i \epsilon_k t} e^{-i p_k x} \ket{p_k,\sigma}_S 
\end{equation}
leading to $\ket{\psi_{\pm}(y-x,t)}_{\sigma} = \braket{y|\psi_{\pm}(x,t)}_S = \sum_{\sigma=1}^{2} \sum_{k=-N_S}^{N_S} c^{(\sigma)}_{\pm,k} e^{\mp i \epsilon_k t} e^{i p_k (y-x)}\ket{\sigma}_{\sigma}$, 
\smallskip
which allows us to finally obtain
\begin{equation}\label{statoDirac1}
	\ket{\psi_{\pm}(\xi,t)}_{\sigma} = \sum_{\sigma=1}^{2} \sum_{k=-N_S}^{N_S} c^{(\sigma)}_{\pm ,k} e^{i ( p_k\xi \mp \epsilon_k t )}\ket{\sigma}_{\sigma} \,.
\end{equation}
This state can be expressed through the notation introduced in (\ref{statovett}), namely:
\begin{equation}
	\ket{\psi_{\pm}(\xi,t)}_{\sigma}\equiv	\psi_{\pm}(\xi,t) = \begin{pmatrix}
		\psi_{\pm,1}(\xi,t) \\
		\psi_{\pm,2}(\xi,t)
	\end{pmatrix}
\end{equation}
where the two components $\psi_{\pm,1}(\xi,t)$ and $\psi_{\pm,2}(\xi,t)$ of the spinor are given by
\begin{equation}
	\left\{
	\begin{aligned}
		& \psi_{\pm,1}(\xi,t) = \sum_{k=-N_S}^{N_S} c^{(1)}_{\pm ,k} e^{\mp i\epsilon_k t } e^{ i p_k\xi }  \\
		& \psi_{\pm,2}(\xi,t) = \sum_{k=-N_S}^{N_S} c^{(2)}_{\pm ,k} e^{\mp i\epsilon_k t } e^{ i p_k\xi } \, .
	\end{aligned}
	\right.
\end{equation}

We now turn to the analysis of the coefficients \( c^{(\sigma)}_{\pm ,k} \). 
As in the Klein-Gordon case, we focus on the positive-energy solution, providing only the final result for the negative-energy solution. 
We start by writing 
\begin{equation}\label{a}
	\psi_{+}(\xi,t) = \sum_{k=-N_S}^{N_S} b_{k} u_{+,k}(\xi,t)
\end{equation}
where 
\begin{equation}\label{b}
	u_{+,k}(\xi,t) = \begin{pmatrix}
		u^{(1)}_{+,k}(\xi,t) \\
		u^{(2)}_{+,k}(\xi,t) 
	\end{pmatrix} 
\end{equation}
and
\begin{equation}\label{c}
	\left\{
	\begin{aligned}
		& u^{(\sigma)}_{+,k}(\xi,t) =  \phi_{\sigma,k}(\xi) e^{- i \epsilon_k t }  \\
		& \phi_{\sigma,k}(\xi) = \chi_{\sigma,k}e^{ i p_k\xi } \\
		&  	c^{(\sigma)}_{+,k}= b_k \chi_{\sigma,k} \,.
	\end{aligned}
	\right.
    \vspace{0.2cm}
\end{equation}
We insert $u_{+,k}(\xi,t)$ into equation (\ref{eqDirac}) (or equivalently (\ref{eqDirac2})), thus obtaining:
\begin{equation}
	\left\{
	\begin{aligned}
		&  \epsilon_k \phi_{1,k}(\xi) = -i \frac{\partial}{\partial \xi} \phi_{2,k}(\xi) + m \phi_{1,k}(\xi)\\
		&  \epsilon_k \phi_{2,k}(\xi) = -i \frac{\partial}{\partial \xi} \phi_{1,k}(\xi) - m \phi_{2,k}(\xi)  \,.
	\end{aligned}
	\right.
\end{equation}
We now make use of the fact that 
$\phi_{\sigma,k}(\xi)=\chi_{\sigma,k}e^{ i p_k\xi }$, leading to
\begin{equation}
	\left\{
	\begin{aligned}
		&  (\epsilon_k -m)\chi_{1,k} - p_k \chi_{2,k}=0\\
		& -p_k \chi_{1,k} + (\epsilon_k-m)  \chi_{2,k} = 0 \ ,
	\end{aligned}
	\right. 
\end{equation}
from which we obtain:
\begin{equation}\label{keychi}
	\chi_{2,k}  = \frac{\epsilon_k - m}{p_k} \chi_{1,k}= \frac{p_k}{\epsilon_k + m} \chi_{1,k} \,.
\end{equation}

In order to determine the expression for \( \chi_{1,k} \) (and consequently for \( \chi_{2,k} \)), we adopt the normalization condition for the time component of the conserved four-current \cite{peskin}: 
\begin{equation}
	\begin{split}
		\rho_+ &= \bar{\psi}_{+}(\xi,t) \gamma^0 \psi_{+}(\xi,t) 
        = \psi^{\dagger}_{+}(\xi,t)\psi_{+}(\xi,t) 
		\\& = \psi^{*}_{+,1}(\xi,t)\psi_{+,1}(\xi,t) + \psi^{*}_{+,2}(\xi,t)\psi_{+,2}(\xi,t) \,.
	\end{split}
\end{equation}
where we used $\bar{\psi}_{+}(\xi,t) =  \psi^{\dagger}_{+}(\xi,t)\gamma^0$ and $\left(\gamma^0\right)^2 = \mathbb{1}$.
Using (\ref{a}), (\ref{b}) and (\ref{c}), we thus require:
\begin{widetext}
	\begin{multline}\label{condition}
		\int_{-L/2}^{L/2} d\xi\: \rho_+ =\int_{-L/2}^{L/2} d\xi\:  \psi^{\dagger}_{+}(\xi,t)\psi_{+}(\xi,t) =\int_{-L/2}^{L/2} d\xi\: \left( \psi^{*}_{+,1}(\xi,t)\psi_{+,1}(\xi,t) + \psi^{*}_{+,2}(\xi,t)\psi_{+,2}(\xi,t) \right) \\
		=	\int_{-L/2}^{L/2} d\xi\: \sum_{k,n=-N_S}^{N_S} b^{*}_k b_n \chi^{*}_{1,k} \chi_{1,n}e^{ i \xi(p_n-p_k) } e^{-it(\epsilon_n - \epsilon_k)}   + \int_{-L/2}^{L/2} d\xi\: \sum_{k,n=-N_S}^{N_S}  b^{*}_k b_n \chi^{*}_{2,k} \chi_{2,n}e^{ i \xi(p_n-p_k) } e^{-it(\epsilon_n - \epsilon_k)} \\
		=	L\sum_{k=-N_S}^{N_S} |b_k|^2| \chi_{1,k}|^2 + L\sum_{k=-N_S}^{N_S} |b_k|^2|\chi_{2,k}|^2 = \sum_{k=-N_S}^{N_S} |b_k|^2 \left(L|\chi_{1,k}|^2 + L|\chi_{2,k}|^2\right)=  1 \,.
	\end{multline}
\end{widetext}
Employing the normalization $1= \sum_{k=-N_S}^{N_S} |b_k|^2$, we can rewrite (\ref{condition}) as:
\begin{equation}
	\sum_{k=-N_S}^{N_S} |b_k|^2 \left(  L|\chi_{1,k}|^2 + 2L|\chi_{2,k}|^2 - 1\right)=0
\end{equation}
which, together with (\ref{keychi}), leads to
\begin{equation}
	|\chi_{1,k}|^2 = \frac{1}{L} \left[ 1 + \left(\frac{\epsilon_k - m}{p_k}\right)^2 \right]^{-1}
\end{equation}
and finally to
\begin{equation}
	\chi_{1,k} = \frac{1}{\sqrt{L}} \frac{\sqrt{\epsilon_k + m}}{\sqrt{2\epsilon_k}} \, ,
\end{equation}
where we used $p^2_k = \epsilon_k^2 - m^2$. Consequently, through (\ref{keychi}), we also obtain:
\begin{equation}
	\chi_{2,k} = \frac{1}{\sqrt{L}}  \frac{ \text{sign}(p_k)\sqrt{\epsilon_k - m}}{\sqrt{2 \epsilon_k}} \,.
\end{equation}

Combining all the results above, the general form of the $u_{+,k}(\xi,t)$ reads:
\begin{equation}
	u_{+,k}(\xi,t) = \frac{1}{\sqrt{L}} u_{+,k}  e^{i(p_k\xi - \epsilon_k t)}  \, ,
\end{equation}
where we have defined
\begin{equation}\label{lalla1}
	u_{+,k} = \frac{1}{\sqrt{2\epsilon_k}}\begin{pmatrix}
		\sqrt{\epsilon_k + m} \\  \text{sign}(p_k)\sqrt{\epsilon_k - m}
	\end{pmatrix} \, .
\end{equation}
In an entirely analogous way, one obtains the negative-energy solution:
\begin{equation}
	u_{-,k}(\xi,t) = \frac{1}{\sqrt{L}} u_{-,k} e^{i(p_k\xi + \epsilon_k t)} 
\end{equation}
with
\begin{equation}\label{lalla2}
	u_{-,k} = \frac{1}{\sqrt{2\epsilon_k}}\begin{pmatrix}
		- \sqrt{\epsilon_k - m} \\  \text{sign}(p_k)\sqrt{\epsilon_k + m}
	\end{pmatrix} \,.
\end{equation}
The solutions derived above lead to the well-known positive/negative-energy spinor solutions of the Dirac equation in \(1+1\) dimensions, in agreement with standard treatments. This confirms that the full relativistic dynamics of a spin-\(1/2\) particle emerges consistently within our constraint-based framework.

As in the Klein–Gordon case, we also take here a preliminary step towards second quantization, which will be useful in the next Section. We start again by assuming $R$,$S$ with $N_R,N_S\to\infty$ and $\hat{H}_C$ with continuous unbouned spectrum. Within this framework we define:
\begin{equation}
	u_k(\xi,t) \equiv u_{+,k}(\xi,t)
\end{equation}
and
\begin{equation}
	v_k(\xi,t) \equiv u_{-,-k}(\xi,t)\, .
\end{equation}
The solutions of the Dirac equation (\ref{eqDirac}) (or equivalently (\ref{eqDirac2})) can thus be written:
\begin{equation}
	\psi_{+}(\xi,t) = \sum_{k=-\infty}^{\infty} b_k u_k(\xi,t)
\end{equation}
and
\begin{equation}
	\psi_{-}(\xi,t) = \sum_{k=-\infty}^{\infty} d^{*}_k v_k(\xi,t)
\end{equation}
where
\begin{equation}\label{uvxitiBUONI}
	\left\{
	\begin{aligned}
		&  u_k(\xi,t) = \frac{1}{\sqrt{L}} u_{k}  e^{i(p_k\xi - \epsilon_k t)}  \\
		& v_k(\xi,t)  =  \frac{1}{\sqrt{L}} v_{k} e^{- i(p_k\xi - \epsilon_k t)} 
	\end{aligned}
	\right.
\end{equation}
with
\begin{equation}\label{uvBUONI}
	\left\{
	\begin{aligned}
		&  u_k =  \frac{1}{\sqrt{2\epsilon_k}} \begin{pmatrix}
			\sqrt{\epsilon_k + m} \\  \text{sign}(p_k)\sqrt{\epsilon_k - m}
		\end{pmatrix} \\
		& v_k  = \frac{1}{\sqrt{2\epsilon_k}} \begin{pmatrix} 
			- \sqrt{\epsilon_k - m} \\ - \text{sign}(p_k)\sqrt{\epsilon_k + m}
		\end{pmatrix} \, .
	\end{aligned} 
	\right. 
\end{equation}

We emphasize that $u_k(\xi,t)$, $v_k(\xi,t)$ and \( u_k \), \( v_k \), as defined in equations (\ref{uvxitiBUONI}) and (\ref{uvBUONI}), satisfy the properties typically required in the mode analysis of Dirac spinors.
Specifically, we have $u^{\dagger}_k u_k  = v^{\dagger}_k v_k = 1$, leading to
\begin{equation}\label{110}
	\begin{aligned}
		& \int_{-L/2}^{L/2} d\xi\: u^{\dagger}_k(\xi,t)u_{n}(\xi,t)=\int_{-L/2}^{L/2} d\xi\: v^{\dagger}_k(\xi,t)v_{n}(\xi,t)={\delta_{k,n}}  \\
		& \sum_{k=-\infty}^{\infty}  u^{\dagger}_k(\xi,t)u_{k}(\xi',t)= \sum_{k=-\infty}^{\infty} v^{\dagger}_k(\xi,t)v_{k}(\xi',t) = \delta(\xi-\xi') 
	\end{aligned}
\end{equation}
and, in addition, the following hold:
\begin{itemize}
	\item $u_k(\xi,t)$, $v_k(\xi,t)$ are eigenvectors of $\hat{H}_D = -i\sigma_1 \frac{\partial}{\partial \xi} + m\sigma_3$ with eigenvalues $\epsilon_k$ and $- \epsilon_k$, respectively;
	\item $\bar{u}_k v_k = u^{\dagger}_k \gamma^0 v_k = 0, \quad \bar{v}_k u_k = v^{\dagger}_k \gamma^0 u_k = 0$; 
	\item $\bar{u}_k u_k =\frac{\epsilon_k}{m}, \quad  \bar{v}_k v_k = - \frac{\epsilon_k}{m}$;
	\item $u^{\dagger}_k v_{-k} =  v^{\dagger}_k u_{-k} =0$;
	\item $u_k u^{\dagger}_k + v_{-k} v^{\dagger}_{-k} = \mathbb{1}$.
\end{itemize}
The relations derived above—including the explicit form of the spinors  $u_k(\xi,t)$, $v_k(\xi,t)$—capture the essential algebraic structure required for the second quantization of the Dirac field and will be used in the next Section.

\section{Second quantization formalism}
We discuss here how a second-quantized formulation can be implemented within our framework. 
	Although the underlying model describes a closed Universe with a finite number of subsystems, we introduce this formalism as a representational tool. While it naturally accommodates multi-particle states, we restrict the analysis to the single-excitation sector, so that both $R$ and $S$ are effectively described as single particles. The field-theoretic structure is thus used only to maintain contact with standard quantum field theory, without addressing genuinely many-particle dynamics. 

Before introducing second-quantized fields on the relational coordinate \( \xi = y - x \), it is however important to clearly state the key assumptions that ensure the mathematical and physical consistency of the formalism.

As anticipated, we assume here that \( \hat{P}_R \) and \( \hat{P}_S \) have unbounded discrete spectra with eigenvalues \( p_k = \frac{2\pi}{L}k \) and \( k \in \mathbb{Z} \). While much of the paper were concerned with wave functions supported over a finite range of momenta (which is physically reasonable for states with finite energy), 
the construction of a field operator in position space requires the full momentum spectrum. If the momentum basis is truncated, the equal-time commutator between the field operator (to be introduced) and its Hermitian conjugate—or, in the Klein–Gordon case, its conjugate momentum—does not vanish for \( \xi \ne \xi' \), spoiling the locality structure. 
By allowing \( \hat{P}_R \) and \( \hat{P}_S \) to span all \( k \in \mathbb{Z} \), we recover the key property:
\begin{equation}\label{discordia}
	\sum_{k=-\infty}^{\infty} e^{i p_k(\xi - \xi')} = L \delta(\xi - \xi') \quad \text{for } \xi, \xi' \in \left[-\frac{L}{2}, \frac{L}{2}\right) \, ,
\end{equation}
which ensures that the fields defined below possess the correct commutation (or anticommutation) relations in position space. Importantly, this extension remains compatible with the energy-neglecting approximation for the reference \( R \), provided that \( M \gg m \) and \( M \gg |p_k| \) for all the momenta relevant to the dynamics.

Finally, we note that, in this framework, we also make use of a continuous and unbounded energy spectrum for $C$, to ensure that the global energy constraint can be exactly satisfied. 
As discussed in Section~II.A, this remains consistent with the hypothesis of a finite Universe by considering that the relevant dynamics are effectively captured within a large but finite energy window.

\subsection{Schrödinger case}
We consider the case of \( R \) and \( S \) as Schrödinger free particles. Although the second quantization of the Schrödinger field is straightforward, we include it here as a useful stepping stone to understand how our relational framework accommodates the field-theoretic formulation.

We start studying the regime in which the kinetic energy of \( R \) can be neglected. The energy constraint reads:
\begin{equation}
	\left(\hat{H}_C + \frac{\hat{P}^2_S}{2m}\right)\ket{\Psi} \approx 0 
\end{equation}
from which, in combination with (\ref{2}), we obtain
\begin{equation}
	\ket{\Psi} = \sum_{k=-\infty}^{\infty} c_k \ket{E=- \frac{p^2_k}{2m}}_C \otimes \ket{p=-p_k}_R\otimes \ket{p_k}_S
\end{equation}
and finally (after projecting onto the bases $\{ \ket{t}_C\}$, $\{ \ket{x}_R \}$, $\{ \ket{y}_S \}$) the wave function
\begin{equation}
	\begin{split}
		\psi(\xi,t) &= \braket{t,x,y|\Psi}
		\\&	= \sum_{k=-\infty}^{\infty} a_k u_k(\xi,t)
		\\& =  \frac{1}{\sqrt{L}} \sum_{k=-\infty}^{\infty} a_k e^{i(p_k\xi - \frac{p^2_k}{2m} t)}
	\end{split}
\end{equation}
where $c_k=\frac{1}{\sqrt{L}} a_k$ and $u_k(\xi,t)= \frac{1}{\sqrt{L}}e^{i(p_k\xi - \frac{p^2_k}{2m} t)}$.

It is thus natural to promote
\begin{equation}
	\psi(\xi,t),\psi^{*}(\xi,t) \longrightarrow \hat{\psi}(\xi,t),\hat{\psi}^{\dagger}(\xi,t) \,,
\end{equation}
where the field operators can be expanded as follows:
\begin{equation}\label{espansionePSI}
	\left\{
	\begin{aligned}
		& \hat{\psi}(\xi,t) = \sum_{k=-\infty}^{\infty} \hat{a}_k u_k(\xi,t) \\ 
		& \hat{\psi}^{\dagger}(\xi,t) = \sum_{k=-\infty}^{\infty} \hat{a}^{\dagger}_k u^{*}_k(\xi,t)  
	\end{aligned}
	\right.
\end{equation}
with $\hat{a}^{\dagger}_k,\hat{a}_k$ the usual creation and annihilation operators.
For $\hat{\psi}(\xi,t)$ and $\hat{\psi}^{\dagger}(\xi,t)$ we require the bosonic commutation relations:
\begin{equation}\label{commutatoriPSI}
	\begin{split}
		& \left[\hat{\psi}(\xi,t),\hat{\psi}^{\dagger}(\xi',t) \right] = \delta(\xi-\xi') \\&
		\left[\hat{\psi}(\xi,t),\hat{\psi}(\xi',t) \right] = \left[\hat{\psi}^{\dagger}(\xi,t),\hat{\psi}^{\dagger}(\xi',t) \right] =0 \,. 
	\end{split}
\end{equation}
Together with \( \int_{-L/2}^{L/2} d\xi\, u^{*}_k(\xi,t) u_n(\xi,t) = \delta_{k,n} \), as a consequence of (\ref{discordia}) we now also have: 
\begin{equation}
	\sum_{k=-\infty}^{\infty} u^{*}_k(\xi,t) u_k(\xi',t) = \delta(\xi-\xi')
\end{equation}
which, combined with equations (\ref{commutatoriPSI}), guarantees that the canonical commutation relations for \( \hat{a}_k \) and \( \hat{a}^\dagger_k \) are satisfied, namely
\begin{equation}
	\begin{split}
		& [\hat{a}_n, \hat{a}^\dagger_{k}] = \delta_{n,k} \\&
		[\hat{a}_n, \hat{a}_{k}] = [\hat{a}^\dagger_n, \hat{a}^\dagger_{k}] = 0 \,. 
	\end{split}
\end{equation}
We can thus define the Hamiltonian density:
\begin{equation}
	\hat{h}_{\text{eff}} =  \hat{\psi}^{\dagger}(\xi,t) \left(- \frac{1}{2m}\frac{\partial^2}{\partial \xi^2}\right)  \hat{\psi}(\xi,t)\,, 
\end{equation}
from which we easily obtain
\begin{equation}\label{HSsecondaquantizz}
	\hat{H}_{\text{eff}} = \int_{-L/2}^{L/2} d\xi \: \hat{h}_{\text{eff}} = \sum_{k=-\infty}^{\infty} \frac{p^2_k}{2m} \hat{a}^{\dagger}_{k}\hat{a}_{k} \,. 
\end{equation}
We stress that, although \( \hat{H}_{\mathrm{eff}} \) governs the effective dynamics of \( S \), it should not be interpreted as the fundamental Hamiltonian of the system. Rather, it emerges from the global constraint and the relational structure of the theory, and thus encapsulates the dynamics of \( S \) relative to the chosen reference subsystem. Furthermore, in the present construction, the effective Hamiltonian preserves the particle number, so that the dynamics within each excitation sector remains decoupled. 

In addition to the Hamiltonian $\hat{H}_{\text{eff}}$, we also find:
\begin{equation}\label{mom}
	\hat{P}_{\text{eff}} = \int_{-L/2}^{L/2} d\xi \:  \hat{\psi}^{\dagger}(\xi,t) \left(- i \frac{\partial}{\partial \xi}\right)  \hat{\psi}(\xi,t) = \sum_{k=-\infty}^{\infty} p_k \hat{a}^{\dagger}_{k}\hat{a}_{k} \,,
\end{equation}
representing the total effective momentum, as emerging from the relational structure of the theory. 

In this framework, together with the momentum states $\ket{p_k} = \hat{a}^{\dagger}_k\ket{0}$, we can define the position states:
\begin{equation}\label{xistate}
	\ket{\xi}=  \hat{\psi}^{\dagger}(\xi)\ket{0} =  \sum_{k=-\infty}^{\infty} u^{*}_k(\xi) \hat{a}^{\dagger}_k \ket{0} 
\end{equation}
where $u_k(\xi)  = \frac{1}{\sqrt{L}}e^{ip_k\xi}$. These states are eigenstates of the Hermitian operator:
\begin{equation}
	\hat{\xi} = \int_{-L/2}^{L/2} d\xi \: \xi \ket{\xi}\bra{\xi}, \quad \xi \in \left[-\frac{L}{2},\frac{L}{2}\right)\,,
\end{equation}
describing the position of the system particle with respect to the reference $R$. 
To recover the time-dependent wave function that solves the Schrödinger equation~(\ref{eqS}), we simply write:
\begin{equation}
	\psi(\xi,t) = \bra{0}\hat{\psi}(\xi,t)\ket{\psi}
\end{equation}
where
\begin{equation}
	\ket{\psi} = \int_{-L/2}^{L/2}d\xi' \: \psi(\xi',0)\hat{\psi}^{\dagger}(\xi')\ket{0}\,.
\end{equation}
In this representation, the relational spatial structure is encoded in the field operators, which depend on the variable \( \xi \), and in the states \( \ket{\xi} \) generated by the action of $\hat{\psi}^{\dagger}(\xi)$ on the vacuum. 

To conclude this Section, we move to the case in which the kinetic energy of the reference subsystem \( R \) cannot be neglected, where the energy constraint reads: 
\begin{equation}
	\left(\hat{H}_C + \frac{\hat{P}^2_R}{2M}+ \frac{\hat{P}^2_S}{2m}\right)\ket{\Psi} = 0\, ,
\end{equation}
from which (together with constraint (\ref{2})) we obtain the wave function
\begin{equation}
	\psi(\xi,t) 
	=  \frac{1}{\sqrt{L}}\sum_{k=-\infty}^{\infty} a_k e^{i(p_k\xi - \frac{p^2_k}{2\mu} t)}
\end{equation}
\vspace{0.2em}
where $\mu= \frac{mM}{m + M}$.
The entire discussion regarding the introduction of the field operators is the same as above and will not be repeated. 
The difference is that $R$ now enters the equation, leading to the Hamiltonian density:
\begin{equation}
	h'_{\mathrm{eff}} = \hat{\psi}^{\dagger}(\xi,t) \left(- \frac{1}{2\mu}\frac{\partial^2}{\partial \xi^2}\right)  \hat{\psi}(\xi,t)\,, 
\end{equation}
from which we obtain
\begin{equation}\label{Heffsecondaquantizz}
	\hat{H}'_{\text{eff}} = \int_{-L/2}^{L/2} d\xi \: \hat{h}'_{\text{eff}} = \sum_{k=-\infty}^{\infty} \frac{p^2_k}{2\mu} \hat{a}^{\dagger}_{k}\hat{a}_{k} \,.
\end{equation}
The momentum is instead found to coincide with (\ref{mom}).
We thus recover the result that the system \( R+S \) behaves as a single particle with the reduced mass \( \mu \). In the limit \( M \gg m \), we find \( \mu \approx m \). Accordingly, we have \( \hat{H}'_{\text{eff}} \approx \hat{H}_{\text{eff}} \), consistently with the assumption that the kinetic energy of \( R \) can be neglected. This confirms that, when the reference becomes very massive, the usual dynamics for the system \( S \) alone is correctly recovered.

\subsection{Klein-Gordon case}
We study here the quantization of the Klein-Gordon field, assuming the system \( S \) to be a spin-\(0\) particle. We neglect the energy of \( R \), and thus we work within the framework introduced at the end of Section IV.

It is important to notice that, in our framework, the total energy constraint is not imposed in quadratic form. As a result, it does not naturally accommodate both positive- and negative-energy solutions within a single equation, as discussed in Section~IV. To consistently include the full set of solutions, we promote the global state \( \ket{\Psi} \) to a two-component object. This allows us to encode the positive- and negative-energy branches separately, and to represent the constraint as a matrix acting on this extended space. Namely, we write:
\begin{equation}
	\ket{\Psi} = \begin{pmatrix}
		\ket{\Psi_{+}} \\ \ket{\Psi_{-}} 
	\end{pmatrix}
\end{equation}
where $\ket{\Psi_{+}}$ and $\ket{\Psi_{-}}$ independently satisfy
\begin{equation}
	\left\{
	\begin{aligned}
		&  \hat{E}_+ \ket{\Psi_{+}} = \left(\hat{H}_C + \sqrt{\hat{P}^2_S + m^2}\right)\ket{\Psi_{+}} \approx 0 \\
		&  \hat{E}_- \ket{\Psi_{-}} = \left(\hat{H}_C - \sqrt{\hat{P}^2_S + m^2}\right)\ket{\Psi_{-}} \approx 0
	\end{aligned} 
	\right. 
\end{equation}
and
\begin{equation}\label{ultimo}
	\left\{
	\begin{aligned}
		&\hat{M}\ket{\Psi_{+}}= \left(\hat{P}_R + \hat{P}_S \right) \ket{\Psi_{+}} = 0 \\
		& \hat{M}\ket{\Psi_{-}}=\left(\hat{P}_R + \hat{P}_S \right) \ket{\Psi_{-}} = 0  \,.
	\end{aligned} 
	\right. 
\end{equation}
The energy and momentum constraints on the global state $\ket{\Psi}$ thus read:
\begin{equation}
	\hat{C}_E \ket{\Psi} = \begin{pmatrix}
		\hat{E}_+ & 0 \\ 0 & \hat{E}_-
	\end{pmatrix} 
	\begin{pmatrix}
		\ket{\Psi_{+}} \\ \ket{\Psi_{-}} 
	\end{pmatrix} \approx 0
\end{equation}
and
\begin{equation}
	\hat{C}_M \ket{\Psi} = \begin{pmatrix}
		\hat{M} & 0 \\ 0 & \hat{M}
	\end{pmatrix} 
	\begin{pmatrix}
		\ket{\Psi_{+}} \\ \ket{\Psi_{-}} 
	\end{pmatrix} = 0 \,,
\end{equation}
which lead to
\begin{equation}
	\left\{
	\begin{aligned}
		&  \ket{\Psi_{+}} = \sum_{k=-\infty}^{\infty} c_{+,k} \ket{E=- \epsilon_k}_C \otimes \ket{p=-p_k}_R\otimes \ket{p_k}_S \\
		&  \ket{\Psi_{-}} =   \sum_{k=-\infty}^{\infty} c_{-,k} \ket{E=+\epsilon_k}_C \otimes \ket{p=-p_k}_R\otimes \ket{p_k}_S
	\end{aligned} 
	\right. 
\end{equation}
where we have reintroduced $\epsilon_k =\sqrt{p^2_k + m^2}$.

In order to extract the global wave function in the usual form, we must now project $\ket{\Psi}$ not only onto the basis states \( \{\ket{t}_C\} \), \( \{\ket{x}_R\} \), and \( \{\ket{y}_S\} \), but also along the equal-weight superposition of the auxiliary components. Namely, we define:
\begin{multline}\label{mamma}
	\psi(\xi,t) \equiv \left[ \bra{t,x,y}\otimes \left(1, 1\right)\right]\ket{\Psi} 
	\\= \sum_{k=-\infty}^{\infty} \left(a_k u_{+,k}(\xi,t) + b^{*}_k u_{-,k}(\xi,t) \right) 
	\\= \frac{1}{\sqrt{L}}\sum_{k=-\infty}^{\infty} \left(\frac{a_k}{\sqrt{2\epsilon_k}} e^{i(p_k\xi - \epsilon_k t)} + \frac{b^{*}_k}{\sqrt{2\epsilon_k}} e^{i(p_k\xi + \epsilon_k t)} \right)
\end{multline}
where 
\begin{equation}
	c_{+,k} =  \frac{1}{\sqrt{L}\sqrt{2\epsilon_k}} a_k \,, \quad c_{-,k}=\frac{1}{\sqrt{L}\sqrt{2\epsilon_k}} b^{*}_k \,,
\end{equation}
$u_{+,k}$, $u_{-,k}$ are defined in (\ref{61}), (\ref{63}), and the vector $(1,1)^{\text{T}}$ selects the wave function as the coherent superposition of the positive- and negative-energy solutions. 

We note that the two-component structure used here does not correspond to an intrinsic degree of freedom. Rather, it serves as a formal device to encode both positive- and negative-energy solutions within a single unified wave function. The projection onto the $(1, 1)^{\text{T}}$ direction simply recovers the total physical solution as a superposition of the two energy branches.

As at the end of Section IV, we can now identify 
\begin{equation}
u_k\equiv u_{+,k}\, , \quad u^{*}_k= u_{-,-k}\, ,
\end{equation}
and rewrite (\ref{mamma}) as
\begin{multline}\label{kg}
	\psi(\xi,t) = \sum_{k=-\infty}^{\infty} \left(a_k u_k(\xi,t) + b^{*}_k u^{*}_k(\xi,t) \right) 
	\\= \frac{1}{\sqrt{L}}\sum_{k=-\infty}^{\infty} \left(\frac{a_k}{\sqrt{2\epsilon_k}} e^{i(p_k\xi - \epsilon_k t)} + \frac{b^{*}_k}{\sqrt{2\epsilon_k}} e^{-i(p_k\xi - \epsilon_k t)} \right) 
\end{multline}
which is a convenient form of the wave function for proceeding with quantization. Notice that, in our framework, we are describing a complex Klein–Gordon field. 

Starting from (\ref{kg}), and analogously to the standard Klein–Gordon theory, we define the conjugate momentum as 
\begin{equation}
	\pi(\xi,t) = \frac{\partial}{\partial t}\psi^{*}(\xi,t) \, ,
\end{equation}
and we promote:
\begin{equation}
	\begin{aligned}
		&\psi(\xi,t),\psi^{*}(\xi,t) \longrightarrow \hat{\psi}(\xi,t),\hat{\psi}^{\dagger}(\xi,t) \\
		&\pi(\xi,t),\pi^{*}(\xi,t) \longrightarrow \hat{\pi}(\xi,t),\hat{\pi}^{\dagger}(\xi,t)
	\end{aligned}
\end{equation}
where 
\begin{equation}\label{kg1}
	\left\{
	\begin{aligned}
		& \hat{\psi}(\xi,t)= \sum_{k=-\infty}^{\infty} \left(\hat{a}_k u_k(\xi,t) + \hat{b}^{\dagger}_k u^{*}_k(\xi,t) \right)   \\
		& \hat{\psi}^{\dagger}(\xi,t)= \sum_{k=-\infty}^{\infty} \left(\hat{a}^{\dagger}_k u^{*}_k(\xi,t) + \hat{b}_k u_k(\xi,t) \right)  
	\end{aligned} 
	\right. 
\end{equation}
and
\begin{equation}\label{kg2}
	\left\{
	\begin{aligned}
		& \hat{\pi}(\xi,t)= i\sum_{k=-\infty}^{\infty} \epsilon_k \left(\hat{a}^{\dagger}_k u^{*}_k(\xi,t) - \hat{b}_k u_k(\xi,t) \right)  \\
		& \hat{\pi}^{\dagger}(\xi,t)=-i \sum_{k=-\infty}^{\infty}\epsilon_k \left(\hat{a}_k u_k(\xi,t) - \hat{b}^{\dagger}_k u^{*}_k(\xi,t) \right) \, .
	\end{aligned} 
	\right. 
\end{equation}
We can thus impose the commutation relations:
\begin{equation}\label{commutatoriKG}
	\begin{split}
		& \left[\hat{\psi}(\xi,t),\hat{\pi}(\xi',t) \right] = i \delta(\xi-\xi') 
		\\& \left[\hat{\psi}(\xi,t),\hat{\psi}(\xi',t) \right] = \left[\hat{\pi}(\xi,t),\hat{\pi}(\xi',t) \right] =0 
	\end{split}
\end{equation}
from which, using the expansions (\ref{kg1}), (\ref{kg2}), together with (\ref{discordia}), naturally follow 
\begin{equation}
	\begin{split}
		& [\hat{a}_n, \hat{a}^\dagger_{k}] =[\hat{b}_n, \hat{b}^\dagger_{k}] =  \delta_{n,k}
		\\& [\hat{a}_n, \hat{a}_{k}] = [\hat{a}^\dagger_n, \hat{a}^\dagger_{k}] =  
		[\hat{b}_n, \hat{b}_{k}] = [\hat{b}^\dagger_n, \hat{b}^\dagger_{k}] = 0 \,,
	\end{split}
\end{equation}
as expected for bosonic creation/annihilation operators.

	As in the Schrödinger case, we now introduce the Hamiltonian density associated with the effective relative dynamics. This allows us to derive the corresponding effective Hamiltonian \( \hat{H}_{\text{eff}} \), which governs the evolution of the system in the relational picture. We have \cite{greinerfields}:
\begin{widetext}
	
	\begin{equation}
		\hat{h}_{\text{eff}}= \hat{\pi}^{\dagger}(\xi,t)\hat{\pi}(\xi,t) + \frac{\partial}{\partial \xi} \hat{\psi}^{\dagger}(\xi,t) \frac{\partial}{\partial \xi} \hat{\psi}(\xi,t) + m^2  \hat{\psi}^{\dagger}(\xi,t)\hat{\psi}(\xi,t) 
	\end{equation}
	and, defining $E_0 = \sum_{k=-\infty}^{\infty}\epsilon_k$, we obtain 
	
	\begin{equation}
			\hat{H}_{\text{eff}} = \int_{-L/2}^{L/2}d\xi\: \hat{h}_{\text{eff}} - E_0 
			= \sum_{k=-\infty}^{\infty} \epsilon_k \left( \hat{a}^{\dagger}_k\hat{a}_k + \hat{b}_k\hat{b}^{\dagger}_k\right) - E_0
			= \sum_{k=-\infty}^{\infty} \epsilon_k \left( \hat{a}^{\dagger}_k\hat{a}_k + \hat{b}^{\dagger}_k\hat{b}_k\right) 
	\end{equation}
	where we used the relation 
	
	\begin{equation}
		\int_{-L/2}^{L/2} d\xi\: e^{i\xi(p_k \mp p_n)} = L\delta_{k,\pm n}\, .
	\end{equation}
The resulting Hamiltonian, with the vacuum energy \( E_0 \) subtracted, takes the standard form of a sum over particle and antiparticle number operators. 

We emphasize that, in the relativistic case, the structure of the field and the symmetry between positive- and negative-energy modes naturally imply the presence of antiparticle degrees of freedom whenever a particle is considered. Nevertheless, within our framework we restrict the analysis to the single-excitation sector, which allows us to describe either a particle or an antiparticle, both associated with the same underlying constraint. 
	
	In a similar way, we find:
	
	\begin{equation}
			\hat{P}_{\text{eff}} = - \int_{-L/2}^{L/2} d\xi\: \left( \hat{\pi}(\xi,t)\frac{\partial}{\partial \xi}\hat{\psi}(\xi,t) + \hat{\pi}^{\dagger}(\xi,t)\frac{\partial}{\partial \xi}\hat{\psi}^{\dagger}(\xi,t) \right)
			= \sum_{k=-\infty}^{\infty} p_k \left( \hat{a}^{\dagger}_k\hat{a}_k + \hat{b}^{\dagger}_k\hat{b}_k \right)
	\end{equation}
	representing the momentum operator, also written as a sum over particle and antiparticle number operators.
	
\end{widetext}

\clearpage
\newpage
\subsection{Dirac case}
We study the quantization of the Dirac field, assuming the system \( S \) to be a spin-\(1/2\) particle. Also in this case we neglect the energy of \( R \), which is taken to be a spin-\(0\)~particle, 
and thus we start from the framework introduced at the end of Section~V. 

In the case of the Dirac equation, the matrix-valued constraint formalism introduced for the Klein–Gordon case could also be adopted, but it is not strictly necessary. Since the Dirac equation is first-order in time, a single global state with support on both \( +\epsilon_k \) and \( -\epsilon_k \) is already sufficient to generate a conditional wave function for the system that includes both time evolutions \( e^{\mp i \epsilon_k t} \). 

Namely, we take the global state of our quantum Universe $\ket{\Psi}$ satisfying the momentum constraint (\ref{2}) and 
\begin{equation}\label{mmm}
	\left( \hat{H}_C + \hat{P}_S \sigma_1 + m\sigma_3 \right)\ket{\Psi} \approx 0 
\end{equation}
representing the single energy constraint, as opposed to the matrix structure we used in the Klein-Gordon case. 

The general form for $\ket{\Psi}$ can be written:
\begin{widetext}
	\begin{equation}\label{sD}
		\ket{\Psi} = \sum_{\sigma=1}^{2}\sum_{k=-\infty}^{\infty} \left( c^{(\sigma)}_{+,k} \ket{E=- \epsilon_k}_C + c^{(\sigma)}_{-,k} \ket{E=+\epsilon_k}_C\right) \otimes \ket{p=-p_k}_R\otimes \ket{p_k,\sigma}_S = \ket{\Psi_{+}} + \ket{\Psi_{-}}
	\end{equation}
	where both
	\begin{equation}\label{coef}
		\left\{
		\begin{aligned}
			&  \ket{\Psi_{+}} = \sum_{\sigma=1}^{2}\sum_{k=-\infty}^{\infty} c^{(\sigma)}_{+,k} \ket{E=- \epsilon_k}_C \otimes \ket{p=-p_k}_R\otimes \ket{p_k,\sigma}_S \\
			&  \ket{\Psi_{-}} =  \sum_{\sigma=1}^{2} \sum_{k=-\infty}^{\infty} c^{(\sigma)}_{-,k} \ket{E=+\epsilon_k}_C \otimes \ket{p=-p_k}_R\otimes \ket{p_k,\sigma}_S
		\end{aligned} 
		\right. 
	\end{equation}
	independently satisfy the constraints. Projecting thus the state (\ref{sD}) onto the basis states \( \{\ket{t}_C\} \), \( \{\ket{x}_R\} \), \( \{\ket{y}_S\} \) and assuming notation (\ref{statovett}), we directly recover:
	\begin{equation}\label{mezzo}
		\begin{split}
			\psi(\xi,t) = \braket{t,x,y|\Psi} 
			&= \sum_{k=-\infty}^{\infty} \left( b_k u_{+,k}(\xi,t) + d^{*}_k u_{-,k}(\xi,t) \right)
			\\&= \frac{1}{\sqrt{L}}\sum_{k=-\infty}^{\infty}  \left( b_k u_{+,k} e^{i(p_k\xi - \epsilon_k t)} + d^{*}_k u_{-,k}e^{i(p_k \xi + \epsilon_k t)} \right)
		\end{split}
	\end{equation}
\end{widetext}
where the coefficients of (\ref{sD}) and (\ref{coef}) are given by
\begin{equation}
	\begin{pmatrix} c^{(1)}_{+,k} \\ c^{(2)}_{+,k} \end{pmatrix} = \frac{b_k}{\sqrt{L}}u_{+,k}, \quad \begin{pmatrix} c^{(1)}_{-,k} \\ c^{(2)}_{-,k} \end{pmatrix} = \frac{d^{*}_k}{\sqrt{L}}u_{-,k}
\end{equation}
with $u_{+,k}$, $u_{-,k}$ defined in (\ref{lalla1}) and (\ref{lalla2}). Then, after the identification $u_k(\xi,t)\equiv u_{+,k}(\xi,t)$ and $v_k(\xi,t)\equiv u_{-,-k}(\xi,t)$, the state (\ref{mezzo}) becomes
\begin{multline}\label{d}
	\psi(\xi,t) = \sum_{k=-\infty}^{\infty} \left( b_k u_{k}(\xi,t) + d^{*}_k v_{k}(\xi,t) \right)
	\\= \frac{1}{\sqrt{L}} \sum_{k=-\infty}^{\infty}  \left( b_k u_k e^{i(p_k\xi - \epsilon_k t)} + d^{*}_k v_ke^{- i(p_k \xi - \epsilon_k t)} \right)
\end{multline}
where $u_k\equiv u_{+,k}$, $v_k\equiv u_{-,-k}$ are defined in (\ref{uvBUONI}) and satisfy $u^{\dagger}_k u_k  = v^{\dagger}_k v_k = 1$, which leads to relations (\ref{110}), together with all the other mathematical properties listed at the end of the previous Section.

Starting from (\ref{d}), we can now promote:
\begin{equation}
	\psi(\xi,t),\psi^{*}(\xi,t) \longrightarrow \hat{\psi}(\xi,t),\hat{\psi}^{\dagger}(\xi,t)
\end{equation}
where $\hat{\psi}(\xi,t)$ and $\hat{\psi}^{\dagger}(\xi,t)$ are given by
\begin{equation}\label{d1}
	\left\{
	\begin{aligned}
		& \hat{\psi}(\xi,t)= \sum_{k=-\infty}^{\infty} \left(\hat{b}_k u_k(\xi,t) + \hat{d}^{\dagger}_k v_k(\xi,t) \right)   \\
		& \hat{\psi}^{\dagger}(\xi,t)= \sum_{k=-\infty}^{\infty} \left(\hat{b}^{\dagger}_k u^{\dagger}_k(\xi,t) + \hat{d}_k v^{\dagger}_k(\xi,t) \right) \, .
	\end{aligned} 
	\right. 
\end{equation}
For these fields operators we can require the fermionic anti-commutation relations:
\begin{equation}\label{commutatoriD}
	\begin{split}
		& \left\{\hat{\psi}_{\alpha}(\xi,t),\hat{\psi}^{\dagger}_{\beta}(\xi',t) \right\} = \delta_{\alpha,\beta} \delta(\xi-\xi')
		\\& \left\{\hat{\psi}_{\alpha}(\xi,t),\hat{\psi}_{\beta}(\xi',t) \right\} = \left\{\hat{\psi}^{\dagger}_{\alpha}(\xi,t),\hat{\psi}^{\dagger}_{\beta}(\xi',t) \right\} =0 
	\end{split}
\end{equation}
where the indices $\alpha$ and $\beta$ denote the spinor components of the field. From (\ref{commutatoriD}), (\ref{discordia}), and using the property $u_k u^{\dagger}_k + v_{-k} v^{\dagger}_{-k} = \mathbb{1}$, one can easily verify that:
\begin{equation}
	\begin{split}
		& \{\hat{a}_n, \hat{a}^\dagger_{k}\} =\{\hat{b}_n, \hat{b}^\dagger_{k}\} =  \delta_{n,k}
		\\& \{\hat{a}_n, \hat{a}_{k}\} = \{\hat{a}^\dagger_n, \hat{a}^\dagger_{k}\} =  
		\{\hat{b}_n, \hat{b}_{k}\} = \{\hat{b}^\dagger_n, \hat{b}^\dagger_{k}\} = 0\,. 
	\end{split}
\end{equation}
We now introduce the Hamiltonian density \cite{greinerfields}:
\begin{equation}
		\hat{h}_{\text{eff}} = \hat{\psi}^{\dagger}(\xi,t)\left(-i \sigma_1 \frac{\partial}{\partial \xi} + m\sigma_3\right)\hat{\psi}(\xi,t)
	\end{equation}
	and, defining $E_0=-\sum_{k=-\infty}^{\infty}\epsilon_k$, we write 
	\begin{equation}
		\hat{H}_{\text{eff}} = \int_{-L/2}^{L/2}d\xi\: \hat{h}_{\text{eff}} - E_0 \,.
	\end{equation}
	Thanks to the fact that $u_k(\xi,t)$ and $v_k(\xi,t)$ are eigenvectors of the operator $\hat{H}_D=-i\sigma_1 \frac{\partial}{\partial \xi} + m\sigma_3$ with eigenvalues $\epsilon_k$ and $- \epsilon_k$, respectively, and using the properties (\ref{110}) and $u^{\dagger}_k v_{-k} =  v^{\dagger}_k u_{-k} =0$, one easily obtains:
	\begin{equation}
		\begin{split}
			\hat{H}_{\text{eff}} &= \sum_{k=-\infty}^{\infty} \epsilon_k \left( \hat{b}^{\dagger}_k\hat{b}_k - \hat{d}_k\hat{d}^{\dagger}_k\right) - E_0 
			\\&  = \sum_{k=-\infty}^{\infty} \epsilon_k \left( \hat{b}^{\dagger}_k\hat{b}_k + \hat{d}^{\dagger}_k\hat{d}_k \right) 
		\end{split} 
	\end{equation}
	where $\hat{H}_{\text{eff}}$ is again written as a sum over particle and antiparticle number operators, just as in the Klein–Gordon case, reflecting the intrinsic symmetry between positive- and negative-energy modes in the relativistic theory. 
	
	Similarly, for the momentum operator we find:
	\begin{equation}
		\begin{split}
			\hat{P}_{\text{eff}} &=\int_{-L/2}^{L/2}d\xi\: \hat{\psi}^{\dagger}(\xi,t)\left(-i  \frac{\partial}{\partial \xi} \right)\hat{\psi}(\xi,t)
			\\& = \sum_{k=-\infty}^{\infty} p_k  \left( \hat{b}^{\dagger}_k\hat{b}_k + \hat{d}^{\dagger}_k\hat{d}_k \right) \,.
		\end{split}
	\end{equation}
	This concludes our formulation of second quantization across the three dynamical cases considered in this work.

	\section{Conclusions}
	
In this work, we have shown how the standard quantum wave equations—the Schrödinger, Klein--Gordon, and Dirac equations—can emerge from global energy and momentum constraints in a fully relational quantum framework. Building on a setting in which both time and space are defined with respect to internal quantum reference frames, we have demonstrated how the dynamics of conditional states reproduces the familiar structure of these equations without assuming an underlying background spacetime.
	
	In particular, we considered a Universe composed of a quantum clock \( C \), a reference particle \( R \), and a system \( S \). For the Schrödinger case, we derived the correct equation both for the system \( S \) alone, under the assumption that the motion of \( R \) can be neglected, and for the composite system \( R+S \) when the kinetic energy of \( R \) is taken into account. We also included an interaction potential depending on the distance between the system and the reference particle, illustrating how such interactions can be naturally incorporated into the formalism. For the Klein-Gordon and Dirac cases, we focused on the regime in which the energy of the reference particle \( R \) can be neglected, and we studied the corresponding equations for the spin-\(0\) and spin-\( 1/2 \) particles \( S \). The~extension to the joint study of \( R+S \) Klein-Gordon and Dirac systems is left for future work, as it involves technical complications that go beyond the scope of the present paper. 
	Finally, we showed how the formalism of second quantization can be naturally implemented within our framework. 
	We found that, for the Klein–Gordon case, a subtle technical peculiarity arises due to the second-order time derivative in the equation, together with our choice of a non-quadratic energy constraint.
	
More generally, our results support the idea that quantum dynamics may arise from global constraints and entanglement, rather than being imposed on a pre-existing spacetime background. In this sense, the present framework provides a concrete realization of a relational and constraint-based approach, and complements existing works that investigate quantum dynamics from similar perspectives~\cite{scalarparticles,dirac,dias1,dias2,dias3,dias4,nuovo1,nuovo2,nuovo3,nuovo32,giacomini,change1,change2,change3,change4,change5,change6,change7,change8,change9,hoehn1,hoehn2,hoehn3}.

Finally, let us comment on the role of the present approach in the broader context of quantum gravity. The formalism developed here is not intended to provide a complete or directly testable theory of quantum gravity. Rather, it should be understood as a simplified and controlled setting in which key conceptual features—such as the emergence of time and space, and the recovery of dynamical laws from global constraints—can be explicitly realized. In this respect, it shares a common perspective with canonical approaches to quantum gravity, where dynamics is encoded in constraint equations and spacetime structure is not fundamental but emergent. While significant challenges remain in extending these ideas to more realistic and fully relativistic scenarios, the present framework may offer useful insight into how a notion of spacetime and quantum dynamics could arise within a background-independent theory.

Future developments (beyond the already mentioned extension to the composite \( R+S \) Klein--Gordon and Dirac systems) could explore more complex scenarios, such as many-body systems or more general classes of quantum dynamics, further clarifying the scope and limitations of the present approach.


	\section*{Acknowledgements}
	The author acknowledges the Project \lq\lq National Quantum Science and Technology Institute – NQSTI\rq\rq\:Spoke 3 \lq\lq Atomic, molecular platform for quantum technologies\rq\rq, and A. Trombettoni for discussions.

	
	\appendix
	\section{Massless Dirac field and chirality}
	In this Appendix, we consider the massless limit of the Dirac field in \(1+1\) dimensions and examine the structure of its positive-energy solutions. Particular emphasis is placed on the emergence of chiral components and their physical interpretation within our relational framework, where the relevant coordinate is the relative position \( \xi = y - x \), and the system effectively acquires a circular geometry (as the wave function repeats cyclically under translations by integer multiples of $L$).
	
	In the massless case, the energy constraint (\ref{enDirac}), reduces to:
	\begin{equation}
		\left( \hat{H}_C + \hat{P}_S \sigma_1 \right)\ket{\Psi} \approx 0 
	\end{equation}
	which, upon projecting onto the bases \( \{ \ket{t}_C \} \), \( \{ \ket{x}_R \} \), \( \{ \ket{y}_S\} \), and applying the momentum constraint, yields:
	\begin{equation}\label{a2}
		i\frac{\partial}{\partial t} \ket{\psi(\xi,t)}_{\sigma} \approx - i\frac{\partial}{\partial \xi} \sigma_1 \ket{\psi(\xi,t)}_{\sigma} \, .
	\end{equation} 
	Equation (\ref{a2}) shows the massless Dirac equation for the system $S$ expressed in terms of the relative coordinate \( \xi \), retaining the two-component spinor structure labeled~by the index $\sigma$. Using notation (\ref{statovett}) and introducing $\gamma^0 = \sigma_3$ and $\gamma^1 = i\sigma_2$, equation (\ref{a2}) can be recast in the form:
	\begin{equation}\label{a3}
		i \left( \gamma^0 \frac{\partial}{\partial t} +  \gamma^1 \frac{\partial}{\partial \xi} \right) \psi(\xi,t) \approx 0 \,\,.
	\end{equation}
	
	We derive again the solution of equations (\ref{a2}) and (\ref{a3}) directly from the constraints. As the derivation closely mirrors the massive case discussed in the main text, we only highlight the key steps specific to the massless limit. The global state satisfying both the energy and momentum constraints can be written:
	\begin{equation}
		\ket{\Psi_{\pm}} = \sum_{\sigma=1}^{2} \sum_{k=-N_S}^{N_S} c^{(\sigma)}_{\pm,k} \ket{E=\mp |p_k|}_C \ket{p=-p_k}_R \ket{p_k,\sigma}_S 
	\end{equation}
	from which we derive:
	\begin{equation}\label{a5}
		\ket{\psi_{\pm}(\xi,t)}_{\sigma} = \sum_{\sigma=1}^{2} \sum_{k=-N_S}^{N_S} c^{(\sigma)}_{\pm,k} e^{i ( p_k\xi \mp |p_k| t )}\ket{\sigma}_{\sigma} \,.
	\end{equation}
	As metioned, we focus on the positive-energy solution. Using again notation (\ref{statovett}), from (\ref{a5}) we thus obtain:
	\begin{equation}\label{a6}
			\psi_+(\xi,t) = \sum_{\substack{k=-N_S}}^{N_S} b_k u_{k}(\xi,t)
			= \frac{1}{\sqrt{L}}\sum_{\substack{k=-N_S}}^{N_S} b_k u_{k} e^{i(p_k\xi - |p_k|t)}
	\end{equation}
	where $	\begin{pmatrix} c^{(1)}_{+,k} \\ c^{(2)}_{+,k} \end{pmatrix} = \frac{b_k}{\sqrt{L}}u_{k}$ and
	\begin{equation}
		u_{k} =
		\frac{1}{\sqrt{2}}
		\begin{pmatrix}
			1 \\
			\operatorname{sign}(p_k)
		\end{pmatrix} \,. 
	\end{equation}
	We notice that the spinor $u_k$ for $k \ne 0$ is correctly normalized and depends only on the sign of the momentum (the case of $k=0$ will be discussed shortly).
	
	We now examine the chiral structure of these solutions. We define the chirality operator as:
	\begin{equation}
		\gamma^5 \equiv \gamma^0 \gamma^1 = \sigma_1
	\end{equation}
	which has eigenvalues \( \pm 1 \). 
	The operator $\gamma^5$, as defined above, satisfies $\left(\gamma^5\right)^2 = 1$, $\{\gamma^5,\gamma^0\}=0$ and $\{\gamma^5,\gamma^1\}=0$.
	The corresponding chiral projectors are:
	\begin{equation}
		P_R = \frac{1}{2}(\mathbb{1} + \gamma^5) = \frac{1}{2}(\mathbb{1} + \sigma_1), \quad
		P_L = \frac{1}{2}(\mathbb{1} - \gamma^5) = \frac{1}{2}(\mathbb{1} - \sigma_1).
	\end{equation}
	The eigenvectors of \( \gamma^5 \) are:
	\begin{equation}
		\chi_R = \frac{1}{\sqrt{2}}
		\begin{pmatrix}
			1 \\
			1
		\end{pmatrix}, \qquad
		\chi_L = \frac{1}{\sqrt{2}}
		\begin{pmatrix}
			1 \\
			-1
		\end{pmatrix}\,\,,
	\end{equation}
	satisfying $\gamma^5 \chi_{R,L} = \pm \chi_{R,L}$.
	The spinor solutions \( u_k \) of the Dirac equation are therefore given by:
	\begin{equation}
		u_k =
		\begin{cases}
			\chi_R & \text{if } p_k > 0, \\
			\chi_L & \text{if } p_k < 0,
		\end{cases}
	\end{equation}
	where each spinor with $k\ne0$ has definite chirality determined by the sign of the momentum. We notice here that the mode with \( k = 0 \) is not a pure chirality eigenstate. Although it is present in the general field expansion, it does not contribute to the propagation dynamics, as it has zero energy and vanishing group velocity.
	
	With this in mind, we decompose the wave function \( \psi_+(\xi,t) \) in terms of its chiral components:
	\begin{equation}
		\psi_+(\xi, t) = \psi_R(\xi, t) + \psi_L(\xi, t)
	\end{equation}
	with 
	\begin{equation}
		\psi_{R,L}(\xi,t) = P_{R,L} \psi_+(\xi,t)\, .
	\end{equation}
	Inserting this decomposition into the massless Dirac equation (\ref{a3}), yields decoupled evolution equations:
	\begin{equation}
		i\left(\frac{\partial}{\partial t} + \frac{\partial}{\partial \xi}\right)  \psi_R(\xi, t) \approx 0 , \quad
		i \left( \frac{\partial}{\partial t}- \frac{\partial}{\partial \xi} \right) \psi_L(\xi, t) \approx 0\,\,,
	\end{equation}
	so that the chiral components propagate independently:
	\begin{equation}
		\psi_R(\xi, t) = \psi_R(\xi - t), \qquad \psi_L(\xi, t) = \psi_L(\xi + t)\,.
	\end{equation}
	Right-chiral spinors propagate to the right, left-chiral spinors propagate to the left, both with speed $c=1$. 
	
	In our formalism, the wave function \( \psi_+(\xi, t) \) describes the state of the system \( S \) relative to the reference \( R \), with both particles liveing on a circle of length \( L \).  
	Within this geometry, a chiral wave packet does not \lq\lq escape to infinity\rq\rq\:but instead propagates around the \lq\lq relational circle\rq\rq. Nevertheless, each mode has a well-defined group velocity (determined by the sign of $p_k$), corresponding to motion around the circle in a definite direction.  
	Therefore, chirality in this context is naturally interpreted as the \lq\lq sense of rotation\rq\rq\:around the circle: right-chiral spinors (\( \chi_R \), associated with \( p_k > 0 \)) move in one direction (e.g., counterclockwise), while left-chiral spinors (\( \chi_L \), associated with \( p_k < 0 \)) move in the opposite direction (e.g., clockwise).
	This interpretation is fully relational: chirality encodes whether the system rotates clockwise or counterclockwise along the relational circle. It is defined solely from the structure of the relational wave function and does not rely on any external spacetime.



\begin{thebibliography}{100}
		\label{biblio}
		
		
		
		
		\bibitem{dewitt} B. S. DeWitt, \textit{Quantum Theory of Gravity. I. The Canonical Theory}, \href{https://doi.org/10.1103/PhysRev.160.1113}{Phys. Rev. \textbf{160}, 1113 (1967).}
		\bibitem{afundamental} L. Maccone, \textit{A fundamental problem in quantizing general relativity}, \href{https://doi.org/10.1007/s10701-019-00311-w}{Found. Phys. \textbf{49}, 1394–1403 (2019).}	
		\bibitem{QG1} C. Rovelli, \textit{Quantum reference systems}, \href{https://doi.org/10.1088/0264-9381/8/2/012}{Class. Quantum Grav. \textbf{8}, 317 (1991).}
		\bibitem{QG2} C. Rovelli, \textit{Relational quantum mechanics}, \href{https://doi.org/10.1007/BF02302261}{Int. J. Theor. Phys. \textbf{35}, 1637–1678 (1996).}	
		
		\bibitem{isham}  C. J. Isham, \href{https://doi.org/10.1007/978-94-011-1980-1_6}{\textit{Canonical Quantum Gravity and the Problem of Time}, edited by L. A. Ibort, M. A. Rodríguez, \textbf{157} (1993).}
		
		\bibitem{pagewootters} D. N. Page and W. K. Wootters, \textit{Evolution whitout evolution: Dynamics described by stationary observables}, \href{https://doi.org/10.1103/PhysRevD.27.2885}{Phys. Rev. D \textbf{27}, 2885 (1983).}
		\bibitem{wootters} W. K. Wootters, \textit{\lq\lq Time\rq\rq replaced by quantum correlations}, \href{https://doi.org/10.1007/BF02214098}{Int. J. Theor. Phys. \textbf{23}, 701–711 (1984).}
		\bibitem{pagearrow} D. N. Page, \textit{Clock Time and Entropy}. In: Physical Origins of Time Asymmetry, edited by J. J. Halliwell, J. Perez-Mercader, and W. H. Zurek, Cambridge University Press, Cambridge (1993).
		
		
		\bibitem{lloydmaccone} V. Giovannetti, S. Lloyd and L. Maccone, \textit{Quantum time}, \href{https://doi.org/10.1103/PhysRevD.92.045033}{Phys. Rev. D \textbf{92}, 045033 (2015).}
		\bibitem{vedral} C. Marletto and V. Vedral, \textit{Evolution without evolution and without ambiguities}, \href{https://doi.org/10.1103/PhysRevD.95.043510}{Phys. Rev. D \textbf{95}, 043510 (2017).} 
		\bibitem{vedraltemperature} V. Vedral, \textit{Time, (Inverse) Temperature and Cosmological Inflation as Entanglement}. In: \href{https://doi.org/10.1007/978-3-319-68655-4}{R. Renner, S. Stupar, \textit{Time in Physics}, 27-42, Springer (2017).} 
		\bibitem{chapter2} T. Favalli, \textit{Page and Wootters Theory}, in: \textit{On the Emergence of Time and Space in Closed Quantum Systems}, \href{https://doi.org/10.1007/978-3-031-52352-6_2}{Springer Theses, Springer Nature (2024).}
		\bibitem{nostro} T. Favalli and A. Smerzi, \textit{Time Observables in a Timeless Universe}, \href{https://doi.org/10.22331/q-2020-10-29-354}{Quantum \textbf{4}, 354 (2020).} 
		\bibitem{nostro2} T. Favalli and A. Smerzi, \textit{Peaceful coexistence of thermal equilibrium and the emergence of time}, \href{https://doi.org/10.1103/PhysRevD.105.023525}{Phys. Rev. D \textbf{105}, 023525 (2022).}
		\bibitem{nostro3} T. Favalli and A. Smerzi, \textit{A model of quantum spacetime}, \href{https://doi.org/10.1116/5.0107210}{AVS Quantum Sci. \textbf{4}, 044403 (2022).}
		\bibitem{librotommi} T. Favalli, \textit{On the Emergence of Time and Space in Closed Quantum Systems}, \href{https://doi.org/10.1007/978-3-031-52352-6}{edited by Springer Nature (2024).}
		\bibitem{nostro4} T. Favalli and A. Smerzi, \textit{Time Dilation of Quantum Clocks in a Relativistic Gravitational Potential}, \href{https://doi.org/10.3390/e27050489
		}{Entropy 27(5), 489 (2025).}
		\bibitem{interacting} A. R. H. Smith and M. Ahmadi, \textit{Quantizing time: Interacting clocks and systems}, \href{https://doi.org/10.22331/q-2019-07-08-160}{Quantum \textbf{3}, 160 (2019).} 	
		\bibitem{timedilation} A. R. H. Smith and M. Ahmadi, \textit{Quantum clocks observe classical and quantum time dilation}, \href{https://doi.org/10.1038/s41467-020-18264-4}{Nat. Commun. \textbf{11}, 5360 (2020).}	
		\bibitem{simile} A. Boette and R. Rossignoli, \textit{History states of systems and operators}, \href{https://doi.org/10.1103/PhysRevA.98.032108}{Phys. Rev. A \textbf{98}, 032108 (2018).}
		\bibitem{simile2} A. Boette, R. Rossignoli, N. Gigena and M. Cerezo, \textit{System-time entanglement in a discrete time model}, \href{https://doi.org/10.1103/PhysRevA.93.062127}{Phys. Rev. A \textbf{93}, 062127 (2016).} 
		\bibitem{review} P. A. Hoehn, A. R. H. Smith and M. P. E. Lock, \textit{The Trinity of Relational Quantum Dynamics}, \href{https://doi.org/10.1103/PhysRevD.104.066001}{Phys. Rev. D \textbf{104}, 066001 (2021).}
		\bibitem{review2} P. A. Hoehn, A. R. H. Smith and M. P. E. Lock, \textit{Equivalence of approaches to relational quantum dynamics in relativistic settings}, \href{https://doi.org/10.3389/fphy.2021.587083}{Front. Phys. \textbf{9}, 587083 (2021).}
		\bibitem{scalarparticles} N. L. Diaz, J. M. Matera and R. Rossignoli, \textit{History state formalism for scalar particles}, \href{https://doi.org/10.1103/PhysRevD.100.125020}{Phys. Rev. D \textbf{100}, 125020 (2019).}
		\bibitem{dirac} N. L. Diaz and R. Rossignoli, \textit{History state formalism for Dirac's theory}, \href{https://doi.org/10.1103/PhysRevD.99.045008}{Phys. Rev. D \textbf{99}, 045008 (2019).}
		\bibitem{foti} C. Foti, A. Coppo, G. Barni, et al., \textit{Time and classical equations of motion from quantum entanglement via the Page and Wootters mechanism with generalized coherent states}, \href{https://doi.org/10.1038/s41467-021-21782-4}{Nat. Commun. \textbf{12}, 1787 (2021).}
		\bibitem{brukner} E. Castro-Ruiz, F. Giacomini, A. Belenchia and Č. Brukner, \textit{Quantum clocks and the temporal localisability of events in the presence of gravitating quantum systems}, \href{https://doi.org/10.1038/s41467-020-16013-1}{Nat. Commun. \textbf{11}, 2672 (2020).} 
		\bibitem{esp1} E. Moreva, G. Brida, M. Gramegna, V. Giovannetti, L. Maccone and M. Genovese, \textit{Time from quantum entanglement: an experimental illustration}, \href{https://doi.org/10.1103/PhysRevA.89.052122}{Phys. Rev. A \textbf{89}, 052122 (2014).}
		\bibitem{esp2} E. Moreva, M. Gramegna, G. Brida, L. Maccone and M. Genovese, \textit{Quantum time: Experimental multitime correlations}, \href{https://doi.org/10.1103/PhysRevD.96.102005}{Phys. Rev. D \textbf{96}, 102005 (2017).}
		
		
		\bibitem{change1} F. Giacomini, E. Castro-Ruiz and Č. Brukner , \textit{Quantum mechanics and the covariance of physical laws in quantum reference frames}, \href{https://doi.org/10.1038/s41467-018-08155-0}{Nat. Commun. \textbf{10}, 494 (2019).}
		\bibitem{change2} A. Vanrietvelde, P. A. Hoehn, F. Giacomini, E. Castro-Ruiz, \textit{A change of perspective: switching quantum reference frames via a perspective-neutral framework}, \href{https://doi.org/10.22331/q-2020-01-27-225}{Quantum \textbf{4}, 225 (2020).} 
		\bibitem{change3} A. Vanrietvelde, P. A. Hoehn and F. Giacomini, \textit{Switching quantum reference frames in the N-body problem and the absence of global relational perspectives}, \href{https://arxiv.org/abs/1809.05093}{arXiv:1809.05093.}	
		\bibitem{change4} J. M. Yang, \textit{Switching Quantum Reference Frames for Quantum Measurement}, \href{https://doi.org/10.22331/q-2020-06-18-283}{Quantum \textbf{4}, 283 (2020).}	
		\bibitem{change5} F. Giacomini, E. Castro-Ruiz and Č. Brukner, \textit{Relativistic Quantum Reference Frames: The Operational Meaning of Spin}, \href{https://doi.org/10.1103/PhysRevLett.123.090404}{Phys. Rev. Lett. \textbf{123}, 090404 (2019).}
		\bibitem{change6} L. F. Streiter, F. Giacomini and Č. Brukner, \textit{Relativistic Bell Test within Quantum Reference Frames}, \href{https://doi.org/10.1103/PhysRevLett.126.230403}{Phys. Rev. Lett. \textbf{126}, 230403 (2021).}
		\bibitem{change7} AC. de la Hamette and T. D. Galley, \textit{Quantum reference frames for general symmetry groups}, \href{https://doi.org/10.22331/q-2020-11-30-367}{Quantum \textbf{4}, 367 (2020).}
		\bibitem{change8} M. Krumm, P. A. Hoehn and M. P. Mueller, \textit{Quantum reference frame transformations as symmetries and the paradox of the third particle}, \href{https://doi.org/10.22331/q-2021-08-27-530}{Quantum \textbf{5}, 530 (2021).}
		\bibitem{change9} A. Ballesteros, F. Giacomini and G. Gubitosi, \textit{The group structure of dynamical transformations between quantum reference frames}, \href{https://doi.org/10.22331/q-2021-06-08-470}{Quantum \textbf{5}, 470 (2021).}
		
		\bibitem{hoehn1} AC. de la Hamette, T. D. Galley, P. A. Hoehn, L. Loveridge and M. P. Mueller, \textit{Perspective-neutral approach to quantum frame covariance for general symmetry groups}, \href{https://arxiv.org/abs/2110.13824}{arXiv:2110.13824.}
		\bibitem{hoehn2} S. A. Ahmad, T. D. Galley, P. A. Hoehn, M. P. E. Lock and A. R. H. Smith, \textit{Quantum Relativity of Subsystems}, \href{https://doi.org/10.1103/PhysRevLett.128.170401}{Phys. Rev. Lett. \textbf{128}, 170401 (2022).}
		\bibitem{hoehn3} P. A. Hoehn, M. Krumm and M. P. Mueller, \textit{Internal quantum reference frames for finite Abelian groups}, \href{https://doi.org/10.1063/5.0088485}{J. Math. Phys. \textbf{63}, 112207 (2022).}
		
		\bibitem{giacomini} F. Giacomini, \textit{Spacetime Quantum Reference Frames and superpositions of proper times}, 	\href{https://doi.org/10.22331/q-2021-07-22-508}{Quantum \textbf{5}, 508 (2021).}  
		
		\bibitem{nuovo1} M. Suleymanov, I. L. Paiva and E. Cohen, \textit{Nonrelativistic spatiotemporal quantum reference frames}, \href{https://doi.org/10.1103/PhysRevA.109.032205}{Phys. Rev. A \textbf{109}, 032205 (2024).}
		\bibitem{nuovo2} V. Kabel, Č. Brukner and W. Wieland, \textit{Quantum reference frames at the boundary of spacetime}, \href{https://doi.org/10.1103/PhysRevD.108.106022}{Phys. Rev. D \textbf{108}, 106022 (2023).}
		\bibitem{nuovo3} P. A. Hoehn, A. Russo and A. R. H. Smith, \textit{Matter relative to quantum hypersurfaces}, \href{https://doi.org/10.1103/PhysRevD.109.105011}{Phys. Rev. D \textbf{109}, 105011 (2024).}
		
		\bibitem{dias1} N. L. Dias, J. M. Matera and R. Rossignoli, \textit{Spacetime quantum actions}, \href{https://doi.org/10.1103/PhysRevD.103.065011}{Phys. Rev. D \textbf{103}, 065011 (2021).}
		\bibitem{dias2} N. L. Dias, J. M. Matera and R. Rossignoli, \textit{Spacetime quantum and classical mechanics with dynamical foliation}, \href{https://doi.org/10.1103/PhysRevD.109.105008}{Phys. Rev. D \textbf{109}, 105008 (2024).}
		\bibitem{dias3} N. L. Dias, J. M. Matera and R. Rossignoli, \textit{Path integrals from spacetime quantum actions}, \href{https://doi.org/10.1016/j.aop.2025.170052}{Annals of Physics \textbf{479}, 170052 (2025).}
		\bibitem{dias4} N. L. Dias and R. Rossignoli, \textit{Spacetime quantum mechanics for bosonic and fermionic systems}, \href{https://doi.org/10.48550/arXiv.2506.10250}{arXiv:2506.10250.}
		
		\bibitem{nuovo32} A. Singh, \textit{Quantum Space, Quantum Time, and Relativistic Quantum Mechanics}, \href{
			https://doi.org/10.1007/s40509-021-00255-9}{Quantum Stud.: Math. Found. \textbf{9}, 35-53 (2022).}
		
		
		\bibitem{kuchar} K.V. Kuchar, \textit{Time and Interpretations of Quantum Gravity}, \href{https://doi.org/10.1142/S0218271811019347}{Int. J. Mod. Phys. D \textbf{20}, No. supp01, 3-86 (2011).}
		
		\bibitem{pegg} D. T. Pegg, \textit{Complement of the Hamiltonian}, \href{https://doi.org/10.1103/PhysRevA.58.4307}{Phys. Rev. A \textbf{58}, 4307 (1998).}
		\bibitem{chapter3} T. Favalli, \textit{Complement of the Hamiltonian}, in: \textit{On the Emergence of Time and Space in Closed Quantum Systems}, \href{https://doi.org/10.1007/978-3-031-52352-6_3}{Springer Theses, Springer Nature (2024).}
		
		\bibitem{everett} H. Everett, \textit{The Theory of the Universal Wave Function}. In: \href{https://doi.org/10.1515/9781400868056-002}{The Many Worlds Interpretation of Quantum Mechanics, Princeton University Press, Department of Physics, 1-140 (1957).}
		
		\bibitem{peskin} M. E. Peskin and D. V. Schroeder, \textit{An Introduction To Quantum Field Theory}, \href{https://doi.org/10.1201/9780429503559}{edited by CRC Press (1995).}
		\bibitem{greinerrelativistic} W. Greiner, \textit{Relativistic Quantum Mechanics. Wave Equations}, \href{https://doi.org/10.1007/978-3-662-04275-5}{edited by Springer Berlin, Heidelberg (2000).}
		\bibitem{librodirac} B. Thaller, \href{https://doi.org/10.1007/978-3-662-02753-0}{\textit{The Dirac Equation}, edited by Springer-Verlag (1992).}
		\bibitem{greinerfields} W. Greiner, \textit{Field Quantization}, \href{https://doi.org/10.1007/978-3-642-61485-9}{edited by Springer Berlin, Heidelberg (2013).}
		
		
		
		
		
		
		
		
	\end{thebibliography}
\end{document}